\documentclass[twocolumn,showpacs,prl,footinbib,superscriptaddress,amsmath,amssymb,floatfix]{revtex4-1}

\usepackage{graphicx,color}
\usepackage{dcolumn}
\usepackage{bm}
\usepackage{lineno, soul, ulem}
\usepackage{hyperref,url}
\usepackage{xcolor}
\usepackage{array}
\newcommand{\sigmav}[1]{$\langle \sigma v\rangle$ #1}

\newcommand{\planck}[1]{\textit{Planck} #1} 
\newcommand{\healpix}[1]{\texttt{HEALPix} #1}
\newcommand{\galprop}[1]{\texttt{GALPROP} #1}

 \hypersetup{
    colorlinks=true,
    linkcolor=blue,
    filecolor=blue,      
    urlcolor=blue,
    citecolor=blue,
    pdftitle={Dark Matter constraints from Planck observations of the Galactic polarized synchrotron emission - reply to first report}
    }
    
\def \aap  {A\&A}

\def \apj  {ApJ}

\def \jcap  {JCAP}
\def \prd {Phy. Rev. D}

\def \mnras {MNRAS}

\begin{document}

\begin{flushright} 
TTK-22-14
\end{flushright}

\title{Dark Matter constraints from Planck observations  \\ of the Galactic polarized synchrotron emission}

\author{Silvia Manconi}
\email{manconi@physik.rwth-aachen.de}
\affiliation{Institute for Theoretical Particle Physics and Cosmology, RWTH Aachen University, Sommerfeldstr.\ 16, 52056 Aachen, Germany}

\author{Alessandro Cuoco}
\affiliation{Dipartimento di Fisica, Universit\`a di Torino, via P. Giuria, 1, I-10125 Torino, Italy}
\affiliation{Istituto Nazionale di Fisica Nucleare, Sezione di Torino, Via P. Giuria 1, 10125 Torino, Italy}

\author{Julien Lesgourgues}
\affiliation{Institute for Theoretical Particle Physics and Cosmology, RWTH Aachen University, Sommerfeldstr.\ 16, 52056 Aachen, Germany}

\begin{abstract}

Dark Matter (DM) annihilation in our Galaxy may produce a linearly polarized synchrotron signal. We use, for the first time, synchrotron polarization to constrain the DM annihilation cross section by comparing theoretical predictions with the latest polarization maps obtained by the Planck satellite collaboration. We find that synchrotron polarization is typically more constraining than synchrotron intensity by about one order of magnitude, independently of uncertainties in the modeling of electron and positron propagation, or of the Galactic magnetic field. Our bounds compete with Cosmic Microwave Background limits in the case of leptophilic DM.
\end{abstract}

\maketitle

\textbf{\textit{Introduction.}}
High-energetic cosmic-ray (CR) electrons \textit{and} positrons ($e^\pm$ in what follows) can be either accelerated in primary sources such as supernova remnants and pulsar wind nebulae, or produced by spallation of hadronic CRs. Besides, CR $e^\pm$ might also be produced by the annihilation or decay of dark matter (DM) particles in the Galactic DM halo.
Relativistic $e^\pm$ then gyrate and propagate in the interstellar Galactic magnetic field (GMF), and produce secondary emissions such as radio and microwave emission through the synchrotron process. 
The synchrotron signal of DM origin  has been extensively investigated in the past using many radio and microwave surveys, such as WMAP and \textit{Planck}, finding constraints which are complementary to other probes both for the Galactic halo \cite{Blasi:2002ct,Hooper:2008zg,Borriello:2008gy,Regis:2009md,Delahaye:2011jn,2012JCAP...01..005F,Mambrini:2012ue,Bringmann:2014lpa,Egorov:2015eta,Cirelli:2016mrc} and extragalactic targets \cite{Tasitsiomi:2003vw,Colafrancesco:2005ji,siffert10,Fornengo:2011cn,Carlson:2012qc,Regis:2014tga,Hooper:2012jc,Fornengo:2014mna}.
Previous DM searches focused on the  synchrotron total intensity, i.e., the Stokes parameter $I$. However,  synchrotron emission of relativistic $e^\pm$ is partially linearly polarized, and a signal in polarization amplitude (i.e., Stokes $P$) is thus expected. 
We here exploit for the first time the \planck polarization maps in order to constrain Galactic DM signals.
Polarization data have also been used together with total intensity data to study Galactic synchrotron emission and constrain CR propagation  and large scale GMF models in absence of DM annihilation, see e.g. \cite{2010jaffe,Sun:2007mx,strong2011,Bringmann:2011py,Jansson:2012pc,Jansson:2012rt,DiBernardo:2012zu,Mertsch:2013pua,Orlando:2013ysa,2016A&A...596A.103P,2018MNRAS.475.2724O,Jew:2019pyv}.  

The total intensity and the polarization properties of the DM synchrotron emission depend on the strength and orientation of the GMF, as well as on the spatial and energetic distribution of CR $e^\pm$ produced by DM.  
As we shall detail in what follows, the synchrotron intensity and polarization signals are complementary, since they are controlled by different properties of the GMF. We thus expect them to be affected by different systematic uncertainties. 

\textbf{\textit{Microwave maps.}}
The \planck instrument measures both the intensity and polarization of the microwave and sub-millimeter sky, in terms of the Stokes components I (intensity) and Q, U (polarization). The polarization amplitude is defined as $P=\sqrt{Q^2 +U^2}$.
In particular, \planck has so far provided the deepest and highest-resolution view of the microwave and sub-millimeter sky by mapping anisotropies in the cosmic microwave background (CMB) radiation. This made it possible to put strong constraints on the  standard cosmological model and its possible variations \cite{Planck:2018nkj}. 

The \planck sky maps contain contributions from the CMB   
as well as many other astrophysical components ranging from compact Galactic and extragalactic sources to diffuse backgrounds as synchrotron and free-free emission in our Galaxy, see e.g. Fig.4 in Ref.~\cite{Planck:2018nkj}.
Here we are interested in constraining a possible diffuse signal coming from DM annihilation in our Galaxy which, depending on the DM properties, may contribute significantly to the diffuse background. 
Since the CMB contribution is well-measured, we consider CMB-subtracted maps. We refrain from modeling and subtracting any other contribution from  the diffuse backgrounds, such as the Galactic synchrotron emission. We thus derive conservative DM constraints requiring that the DM signal does not exceed the observed emission, once the CMB contribution has been subtracted.

\begin{figure*}[t]
	\centering
\includegraphics[width = 0.49\textwidth]{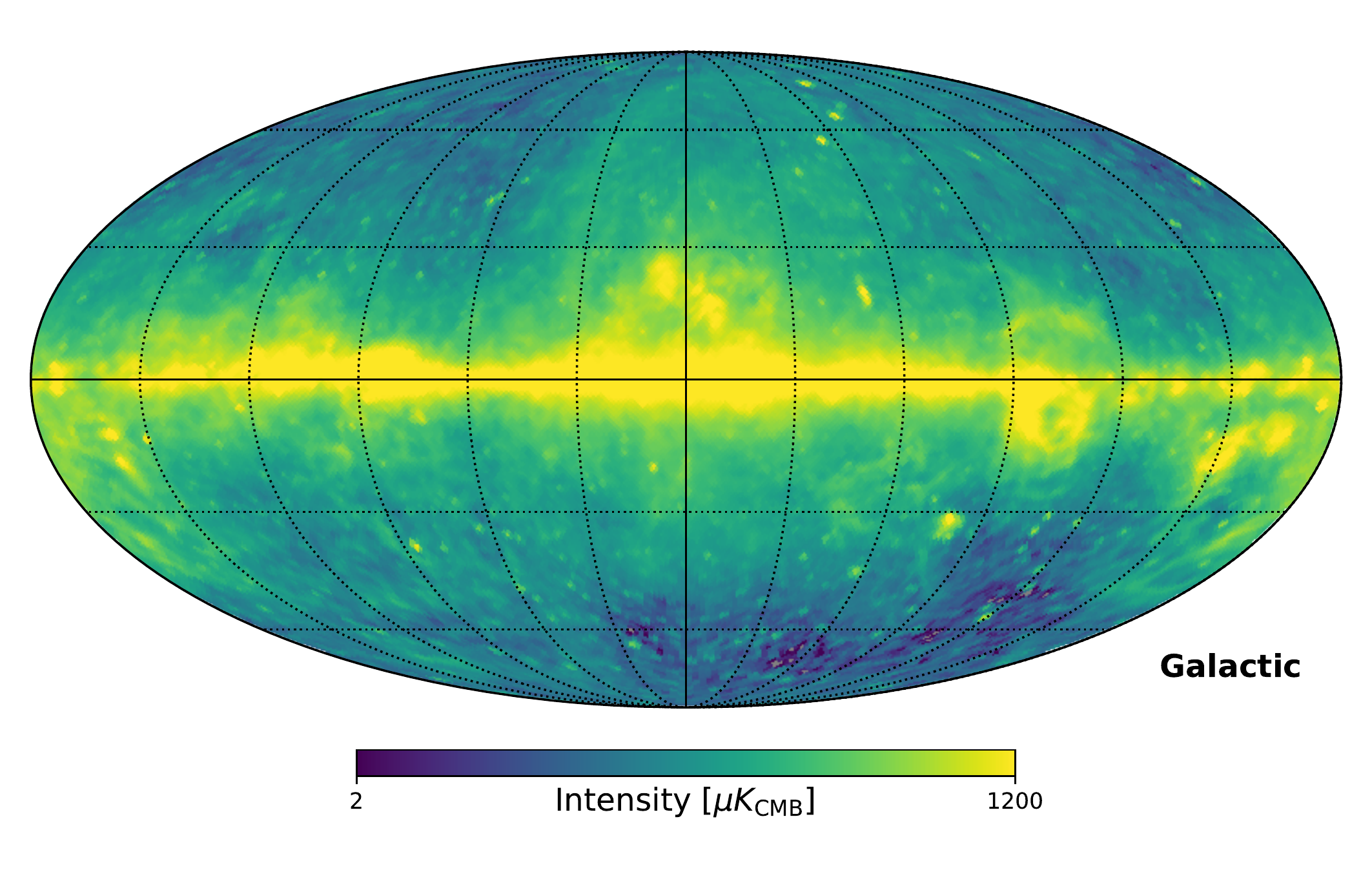}
\includegraphics[width = 0.49\textwidth]{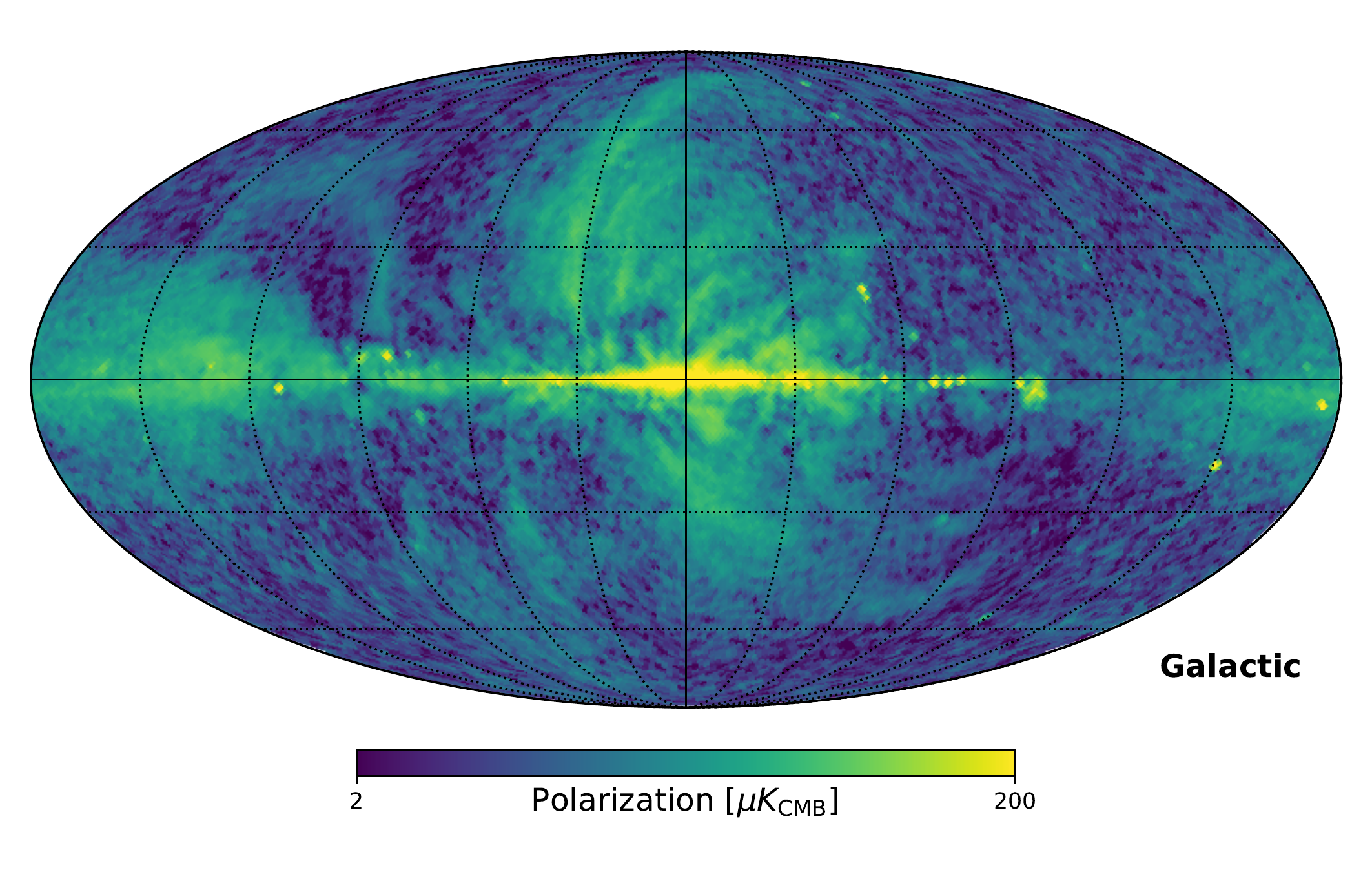}\\
\includegraphics[width = 0.49\textwidth]{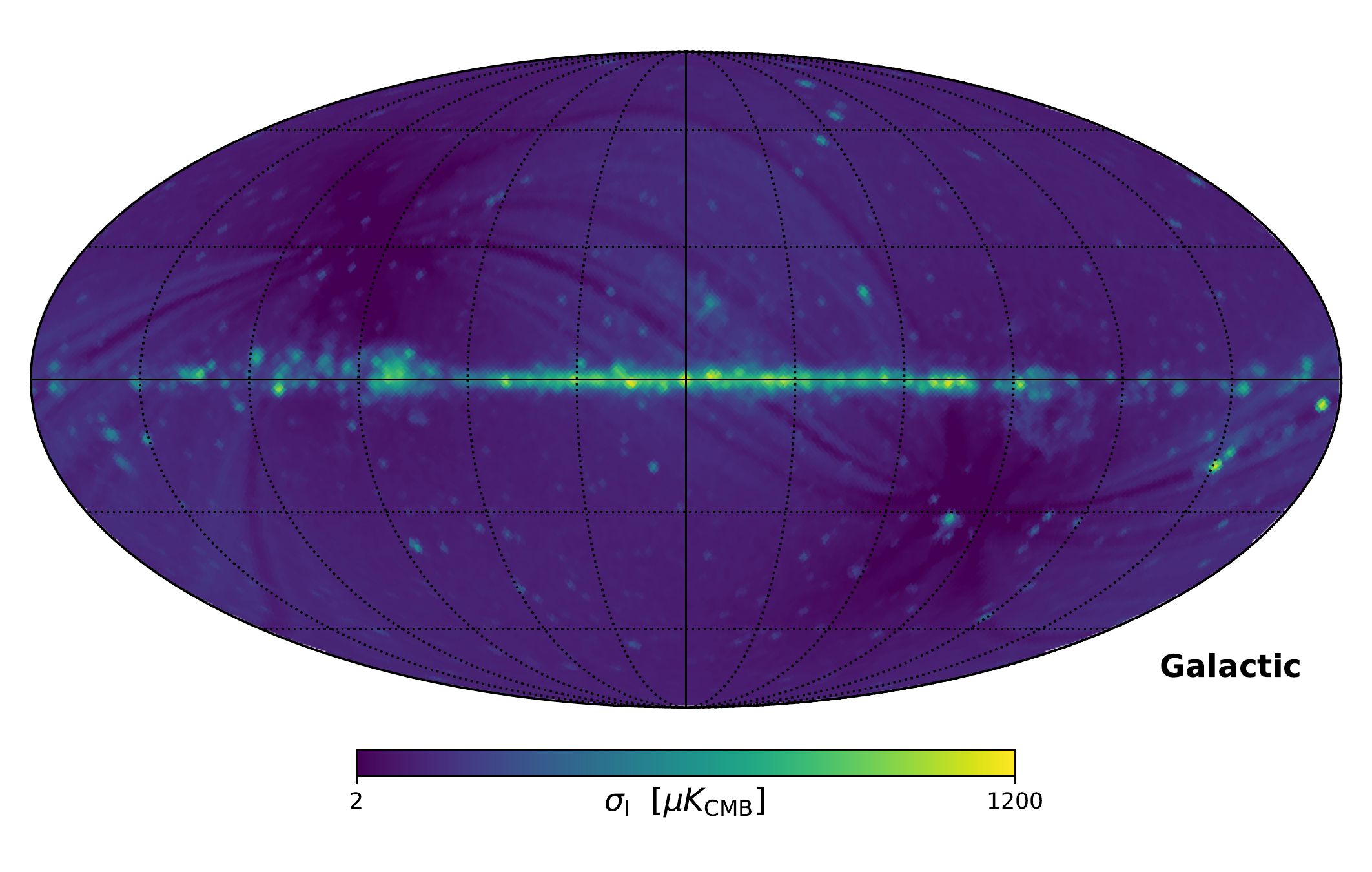}
\includegraphics[width = 0.49\textwidth]{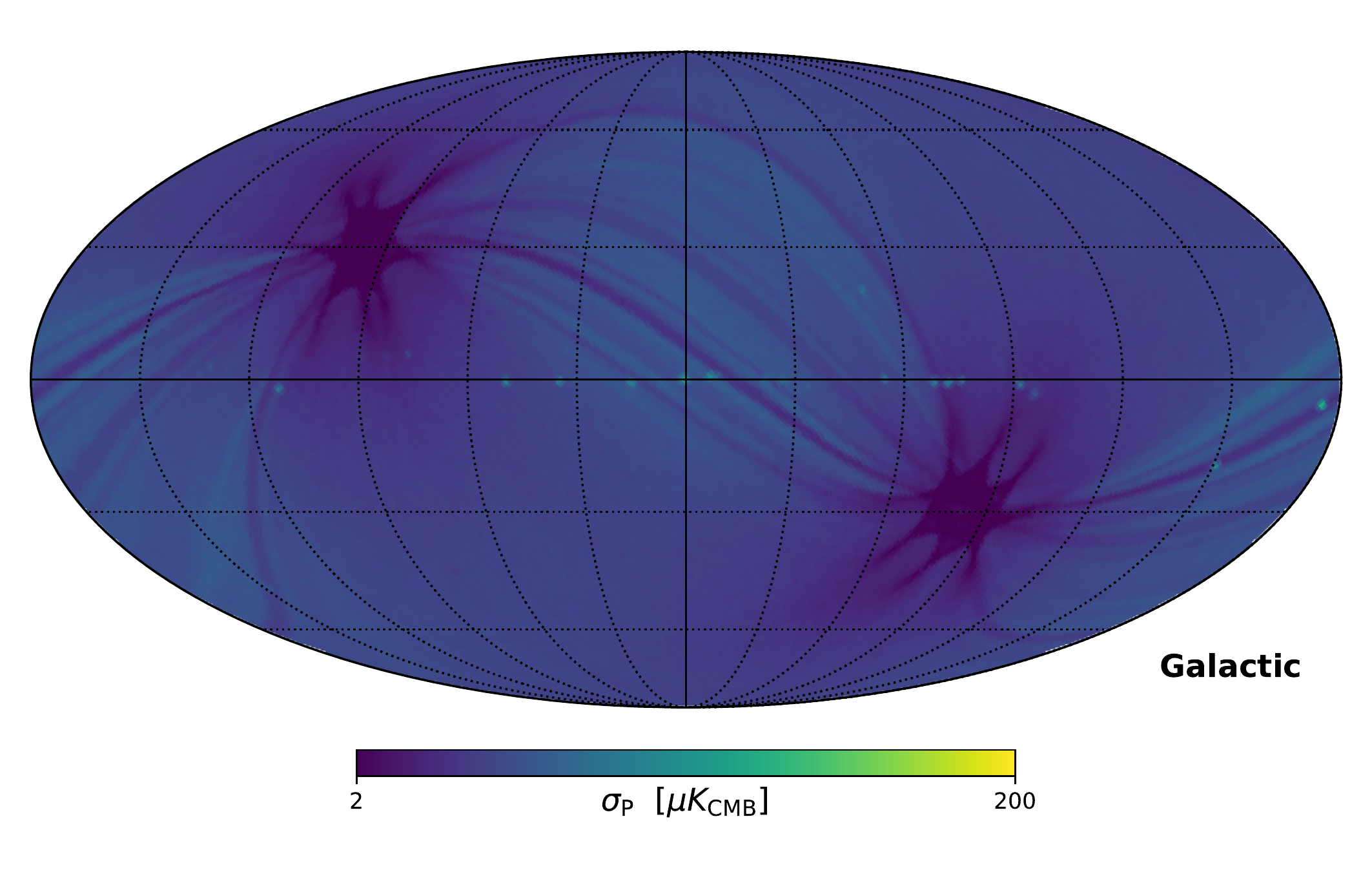}
	\caption{\textbf{\planck 30 GHz data and error estimate ($\sigma$)}. Upper row: NILC CMB-subtracted foreground maps after smoothing with FWHM of $1$ degree and degrading the resolution to NSide=64, for the  Stokes intensity ($I$, left) and polarization ($P$, right).  Lower row: Error estimate for $I$ (left) and  $P$ (right), see text for details. All maps are shown in a Mollweide projection and with a logarithmic color mapping. The minimum and maximum values are manually set to highlight the map structures on top of the noise.	}
	\label{fig:planck}
\end{figure*}

We use data products corresponding to the third release by the \planck collaboration (PR3) for the low frequency instruments (LFI) at $30,44$ and $70$~GHz. Regarding the polarization emission,  the PR3 release supersedes all previous releases thanks to significantly lower contamination from systematic errors \cite{Planck:2018yye}.
CMB-subtracted maps can be obtained using multi-frequency information. We use the maps processed with the NILC method \cite{Planck:2018yye}, which still contain all the diffuse backgrounds.

They can be downloaded from the \planck legacy archive~\footnote{\url{http://pla.esac.esa.int/pla/\#maps}, files named 
LFI\_ CompMap\_Foregrounds-nilc-0XX\_R3.00.fits, 
where XX stands for the frequency channel of 30,44,70 GHz.}
with a resolution of Nside=1024 \footnote{The total number of pixels in the map is related to NSide as $N_{\rm pix}$= 12 Nside$^2$.} in the \healpix pixelization scheme~\cite{Gorski:2004by}. This corresponds to a mean spacing between adjacent pixels of about $0.06$~degrees.
For each frequency, the downloaded files contain three maps, one for each of the three Stokes parameters $I, Q, U$. 
These maps contain the observed blackbody differential brightness temperature 
\footnote{The relation between the brightness temperature and the flux  is recalled in Eq.~(\ref{eq:Tb}), see also  Eq.~(\ref{eq:Jmain}).}
in units of $K_{\rm CMB}$, which is connected to the Rayleigh-Jeans differential brightness temperature in $K_{\rm RJ}$  by a conversion formula that also accounts for color and leakage corrections based on instrument bandpass, see  Ref.~\cite{2016A&A...596A.103P} for more details.

Before comparing the DM predicted $P$ map with observations, we need to build the experimental $P$ map and its error map from the available $Q$ and $U$ maps. This is achieved with the following steps:

(i) \textit{Smoothing}. We first smooth the CMB-subtracted $I,Q,U$ maps with a Gaussian beam of $1$~degree FWHM, in order to increase the signal-to-noise ratio and reduce systematic effects caused by beam asymmetries. 
We then create a polarization amplitude map defined as $P=\sqrt{Q^2 +U^2}$, keeping the original NSide resolution of the $Q,U$ maps. 
The resulting $I$ and  $P$ full-sky maps at $30$~GHz are shown in the upper panels of Fig.~\ref{fig:planck}. 
We provide the full-sky maps at $44,70$~GHz in the supplemental  material \cite{supplemental}. 

(ii) \textit{Error estimation.} 
For the purpose of obtaining robust DM constraints, we need to build error maps from the $I,Q,U$ maps themselves. We estimate the error at each pixel as the variance of all
neighboring pixels up to 0.5 degrees, while sticking to the native NSide resolution.  
This provides an estimate of the noise except in the vicinity of point sources 
\footnote{In the vicinity of sharp features, like point sources, our method is expected to be biased and to return an error estimate larger than the true one.
Nonetheless, this does not affect the final results since pixels with a large error will not contribute to constrain DM.}.
The error map for $P$ is derived from the $Q,U$ error maps using error propagation.
The resulting sky maps for  $\sigma_{\rm I}$ ($\sigma_{\rm P}$) for $30$~GHz are illustrated in the left (right) lower panel of Fig.~\ref{fig:planck}. 
One can see by eye that error maps follow the scanning pattern of \textit{Planck}: the error is smaller where the instrument observes longer, and vice versa.

(iii) \textit{Degrading.}
If needed, the above maps are degraded to a larger pixel resolution.  
While this is straightforward for the $I,P$ maps, for the error maps one needs to take into account that the error scales with the pixel size.
Going from NSide 1024 to a generic lower resolution $l$, we have  $\sigma_{\rm P}^{l}=\sigma_{\rm P}/\sqrt{N_{\rm pix}^{1024}/N_{\rm pix}^{l}}$.
%

\textbf{\textit{Synchrotron from Galactic dark matter.}}
We consider WIMPs as benchmark DM candidates \cite{Roszkowski:2017nbc}, and we concentrate on the annihilation signal.  
However, we stress that the approach presented here could be extended to any search for DM or other exotic particles if they inject $e^\pm$ in the interstellar medium through annihilation and/or decay processes. 

The source term for $e^\pm$ produced from (Majorana fermions) WIMP annihilations in the Milky Way DM halo reads:
\begin{equation}\label{eq:qdm}
 q_{\rm e^\pm} (\mathbf{x}, E) = \frac{1}{2} \left(\frac{\rho_{\rm DM}(\mathbf{x})}{m_{\rm DM}}\right)^2 \Sigma_f \langle \sigma v\rangle_f \frac{dN_{e^{\pm}}^f}{dE}
\end{equation}
where $m_{\rm DM}$ is the DM mass, $\rho_{\rm DM}(\mathbf{x})$ is the DM density profile in the Galaxy (assumed to be spherically symmetric), $f$ runs over the considered DM annihilation channels, \sigmav is the velocity averaged cross section, and ${dN_{e^{\pm}}^f}/{dE}$ is the $e^\pm$ energy spectrum per annihilation for each annihilation channel $f$.

The DM radial distribution $\rho_{\rm DM}(r)$ in the Galaxy at distance $r$  from the halo center can be effectively described by  the  Navarro-Frenk-White (NFW) and generalized NFW density profile~\cite{Navarro:1995iw,Navarro:2008kc},
where we fix the scale radius to $r_S=23$~kpc and enforce the local DM density at the solar position to be $\rho_{\rm DM}(r_\odot =8.5~\rm{kpc})=0.4$~GeV/cm$^3$ \cite{deSalas:2020hbh}. 
To estimate the uncertainties related to the DM radial distribution, particularly relevant for the innermost part, we also consider two additional cases~\cite{supplemental}. 
To avoid numerical divergences at $r \rightarrow 0$ the profiles are truncated as detailed in Ref.~\cite{Egorov:2015eta}. 
Contributions connected to the presence of DM substructures on top of the main, smooth halo could boost the total DM annihilation rate, and  are conservatively not considered here \cite{Ando:2019xlm,Ishiyama:2019hmh}. 

We consider standard WIMPs with masses $m_{\rm DM}$ between $5$~GeV and $1$~TeV annihilating into three representative channels: two leptonic channels,  $\tau^+ \tau^-$ and $\mu^+ \mu^-$, expected to produce more $e^\pm$ in their final states, and one hadronic channel $b\bar{b}$, producing a much softer spectrum. 
The reference thermally averaged annihilation cross section  is $\langle \sigma v\rangle = 3 \times 10^{-26}$ cm$^3$s$^{-1}$. 
The $e^\pm$ energy spectrum ${dN_{e^{\pm}}^f}/{dE}$ for each channel is taken from the PPPC4DMID library \cite{Cirelli:2010xx} and includes electroweak corrections \cite{Ciafaloni:2010ti}.

\textbf{\textit{Cosmic-ray propagation and maps.}}
The propagation of $e^\pm$ in the interstellar medium can be described through a transport equation which can be solved semi-analytically \cite{Maurin:2018rmm} or numerically by different means \cite{Hanasz:2021rfs}.
We here use \texttt{GALPROP} version v54r2766 \footnote{publicy available at \url{https://gitlab.mpcdf.mpg.de/aws/galprop}} as adapted  in Ref.~\cite{Egorov:2015eta} \footnote{publicy available at \url{https://github.com/a-e-egorov/GALPROP_DM}} to numerically solve the transport equation
and predict
the all-sky  synchrotron signal maps from DM annihilations.
In particular, the computation of the total synchrotron intensity and polarization amplitude is based on the  \texttt{GALPROP} developments described in Refs.~\cite{strong2011,Orlando:2013ysa}, and includes free-free absorption, which is however expected to be subdominant at \planck frequencies. 
\texttt{GALPROP} can solve the transport equation both in two and three spatial dimensions. Since the GMFs we consider are intrinsically 3D, the 3D implementation has to be used to obtain correct predictions. We employ  a spatial resolution of $200$~pc in each spatial dimension.  
%

To gauge the uncertainties related to propagation we consider three propagation models taken from the literature \cite{supplemental}.  We employ as a benchmark the plain diffusion model without convection and reacceleration (named \texttt{PDDE}).   Refs.~\cite{2018MNRAS.475.2724O,2019PhRvD..99d3007O} found this model to be in agreement with cosmic-ray, synchrotron and gamma-ray data using a similar \texttt{GALPROP} setup. 
We test also a model with diffusive reacceleration from the same Refs.~\cite{2018MNRAS.475.2724O,2019PhRvD..99d3007O} (named \texttt{DRE}), and a model with convection (named \texttt{BASE}) from the recent Ref.~\cite{Korsmeier:2021brc}.

We note that \texttt{GALPROP} produces synchrotron maps $\mathcal{J}_{I, P}(\nu, b, l)$ in units of energy$^2 \times$flux, i.e., in units of erg cm$^{-2}$/s/Hz/sr, where $\nu$ is the frequency and $b,l$ are galactic coordinates.
We convert this in \textit{brightness temperature} as:
\begin{equation}\label{eq:Tb}
 T_{I, P}(\nu)= \frac{c^2 \mathcal{J}_{I, P}}{2 \nu^2 k_B}\, ,
\end{equation}
which is the temperature that a body with a Rayleigh Jeans (RJ) spectrum would need in order to emit the same intensity at a given frequency $\nu$.
This  defines the RJ brightness temperature in units of Kelvins ($K_{\rm RJ}$).

\textbf{\textit{Magnetic field models.}}
The main systematic uncertainty of the present work is anticipated to be associated to the modeling of the GMF, which is still poorly constrained~\cite{jaffe2019}.
The magnetic field of our Galaxy is know to have at least two components: a large-scale, regular field and an isotropic turbulent, random one. 
The need for an additional component, called 'ordered random' \cite{2010jaffe} or 'striated' \cite{Jansson:2012pc} has been also recently investigated. This new component corresponds to a large scale ordering of the field, and its intensity is expected to be stronger in the regions between the optical spiral arms.
For a comprehensive review on the available tracers, a detailed recap of some current models and their outstanding issues we refer the reader to Ref.~\cite{jaffe2019} (and references therein).
We thus rely on past studies which fitted the most updated GMF models to multiwavelength data. 
To bracket the uncertainties associated to GMF modeling, we consider the following three benchmarks:  
The Sun+10 model proposed in Refs.~\cite{Sun:2007mx,sun2010}, 
the model proposed in Ref.~\cite{2011ApJ...738..192P} (Psh+11), and the more sophisticated model presented by Jansson \& Farrar  for the regular \cite{Jansson:2012pc} and random \cite{Jansson:2012rt} magnetic fields (JF12) \cite{supplemental}.

These models differ both for the regular and  the random MF component. A crucial observation is the fact that intensity and polarization have a different dependence on the MF. While  intensity depends on the total MF (random+ordered), polarization only depends on the regular component. This makes the two probes highly complementary.

\begin{figure}[t]
	\centering
	\includegraphics[width = 0.49\textwidth]{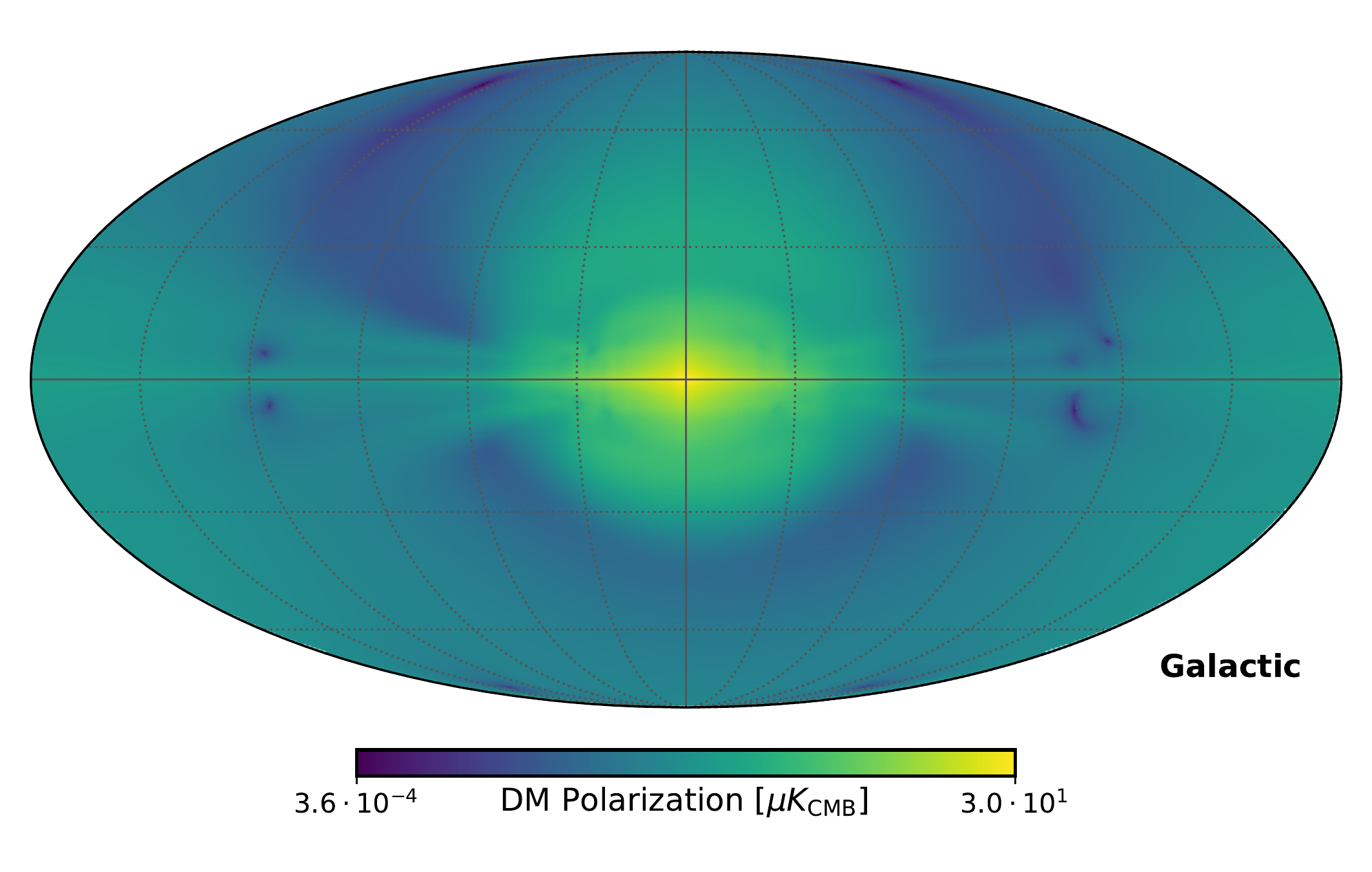}
	\caption{ \textbf{Polarization amplitude of the synchrotron emission from DM at $30$~GHz} in units of $K_{\rm CMB}$ as computed for $m_{\rm DM}=50$~GeV annihilating in $\mu^+ \mu^-$ pairs with a thermal averaged cross section of $\langle \sigma v\rangle = 3 \times 10^{-26}$ cm$^3$s$^{-1}$, and using the PDDE propagation and the Psh+11  GMF model. The sky map is computed for NSide=128 and is shown in Mollview projection. }
	\label{fig:mapPdm}
\end{figure}

\textbf{\textit{Dark matter signal and constraints.}}
 To illustrate the morphology of the polarization DM signals, we show in Fig.~\ref{fig:mapPdm}  the polarization amplitude at $30$~GHz for one GMF model (Psh+11). 
The map is computed for a DM particle of $m_{\rm DM}=50$~GeV annihilating into $\mu^+ \mu^-$ pairs with $\langle \sigma v\rangle = 3 \times 10^{-26}$ cm$^3$s$^{-1}$,  using the PDDE propagation  and for NSide=128. 
The polarization amplitude of the DM signal is, as expected, peaked at the Galactic center and extends away from the plane following the morphology of the regular magnetic field in the Milky Way disk and halo from Psh+11. 

We have validated our results comparing the synchrotron DM maps and spectra with previous works \cite{Egorov:2015eta,Cirelli:2016mrc}, finding similar results when computing the DM signal within the same setup, when possible. 
We refer to \cite{supplemental} for more examples of the intensity and polarization DM signal maps.

In the following we use \planck LFI maps at $30$~GHz as reference, while we show  results using higher frequencies maps in \cite{supplemental}. 
For each simulated DM map, i.e., for each DM mass and annihilation channel, we compute an upper bound on the DM annihilation cross section by requiring that the DM intensity or polarization signal at a given frequency does not exceed the observed \planck signal \textit{plus} the error estimated before, in this way producing limits at the 68\% C.L.
We enforce this requirement in each pixel at $|b|<30$~deg, and we provide the upper limit corresponding to the most constraining pixel.

\begin{figure*}[t]
		\includegraphics[width = 0.49\textwidth]{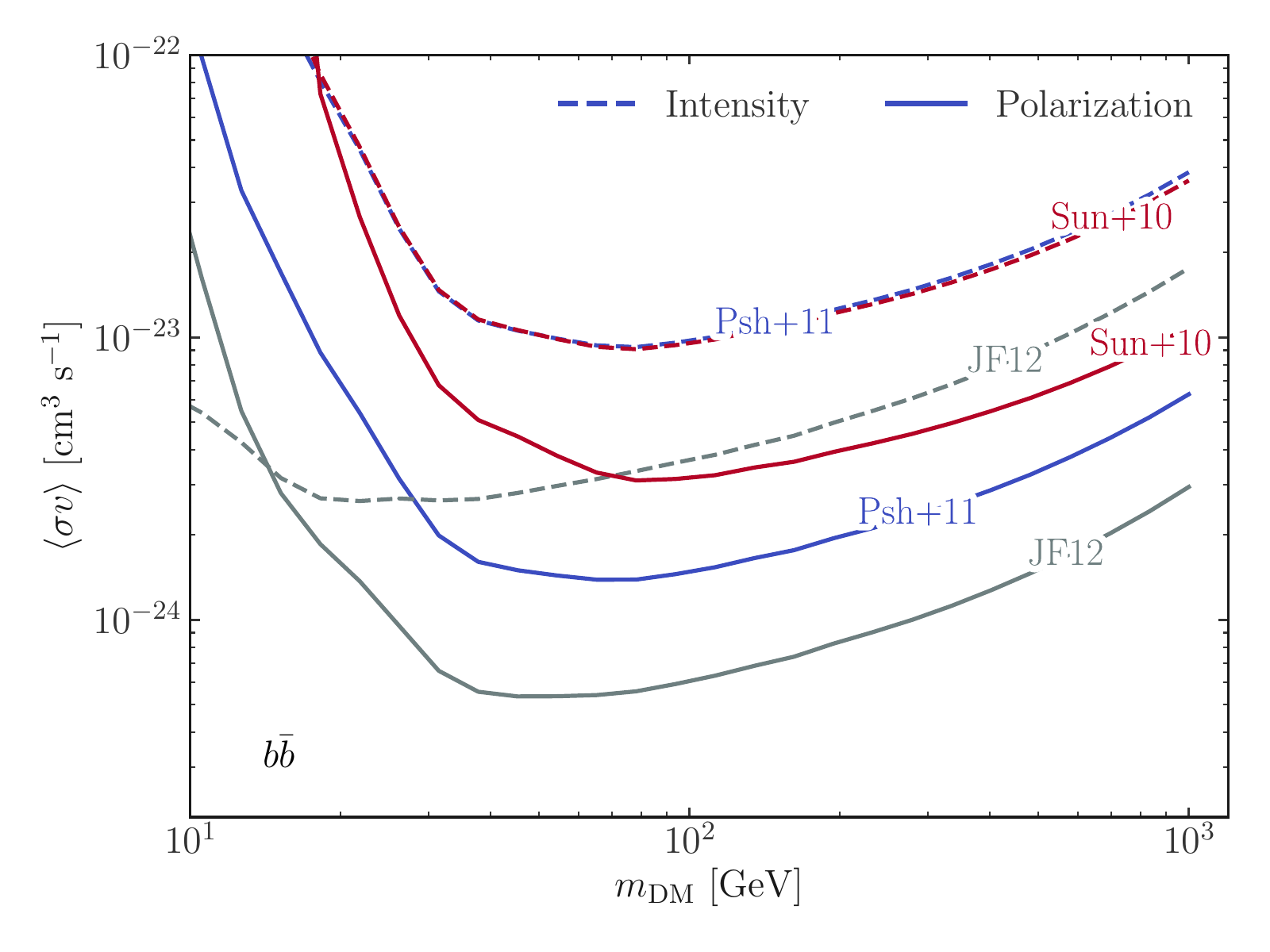}
	\includegraphics[width = 0.49\textwidth]{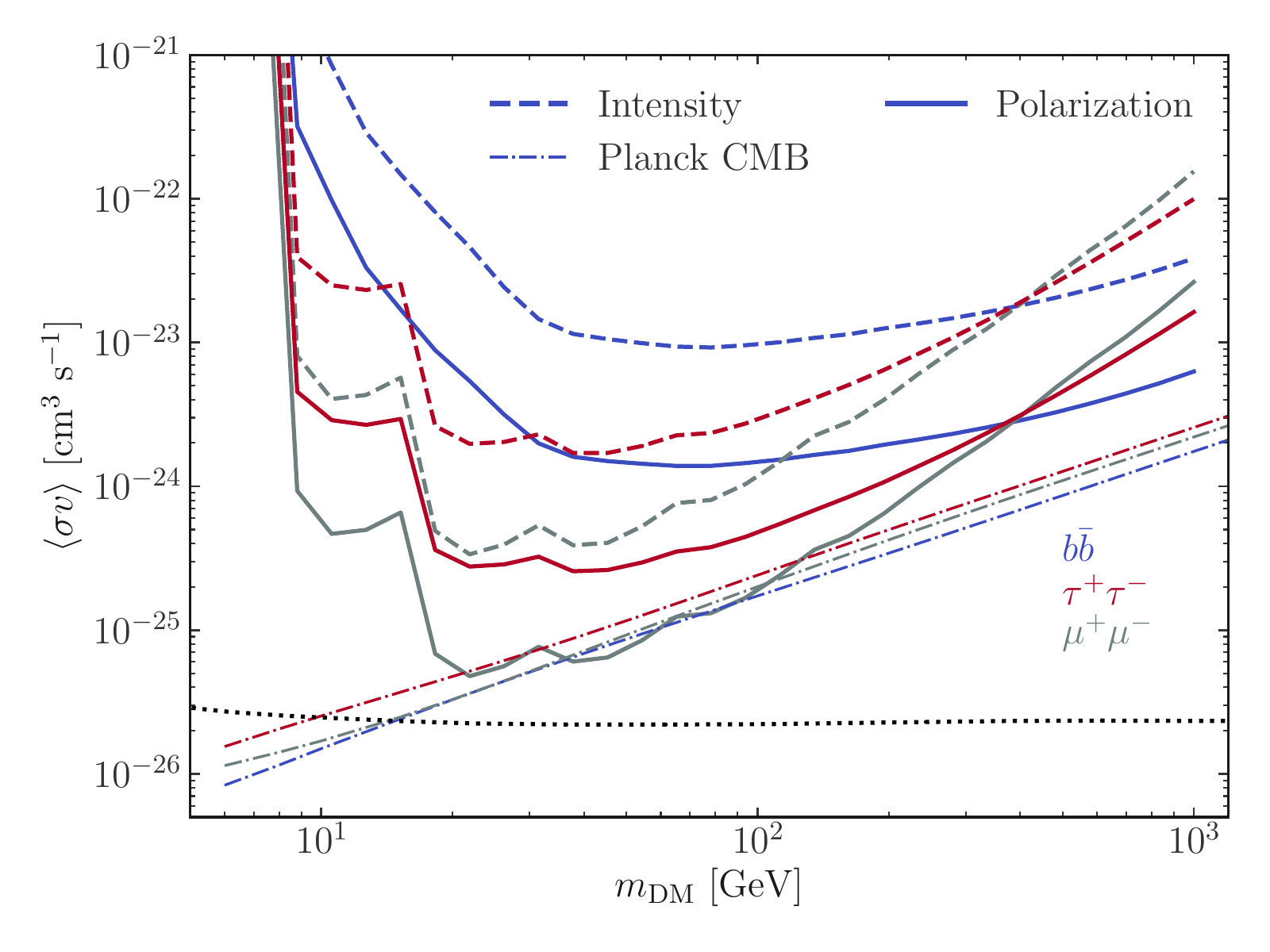}
	\caption{ \textbf{Upper limits on the thermally averaged annihilation cross section as a function of DM mass} as derived from the \planck   intensity (dashed lines) and polarization (solid lines) data at $30$~GHz.
	Left panel: effect of the GMF model for the $\bar{b} b$ channel. Right panel: results for different annihilation channels assuming the Psh+11 GMF model. Results obtained from \planck CMB data \cite{Planck:2018vyg} are reported as dot-dashed lines for comparison. 
	The dotted line indicates the thermal relic cross section \cite{Steigman:2012nb}. Note the different y scales in the two panels.}
	\label{fig:IPul}
\end{figure*}

As a preliminary step, we study the effect of pixel size~\cite{supplemental}. With a small pixel we are sensitive to the detailed morphology of the signal, but the noise per pixel is large, while with a large pixel we have a smaller noise but we lose the details of the morphology.
We find that the constraints are optimized for a choice of an  NSide=128, that we adopt in the following \cite{supplemental}.

Our results for the upper limits obtained using \planck intensity and polarization data are illustrated in Fig.~\ref{fig:IPul} (left) for different GMF models and for the $\bar{b} b$ channel and PDDE propagation setup. 
At fixed GMF model, we find that the polarization maps are more constraining than the intensity maps by almost one order of magnitude for DM masses larger than 20~GeV. 
The Sun+10 and Psh+11 models use the same parametrizations and intensity values for the random field, and thus the intensity constraints are very similar. The random field of the JF12 model has instead a more complicated morphology and a larger strength, which translates into stronger limits by a factor of two.
The different morphology and strength for the ordered GMF translate into an uncertainty of about one order of magnitude in the upper limits obtained with the polarization data. The JF12 model is in this case associated to the most stringent upper limits given the non-zero striated component included.  
We recall that the strength of the GMF is highly degenerate with the normalization of the CR $e^\pm$ density in the Galaxy, and a consistent assessment of the parameters of the GMFs should contextually fit also the CR $e^\pm$ injection and propagation parameters. We leave this assessment to future work, in which potentially stronger constraints can be derived by modeling  and subtracting the astrophysical Galactic synchrotron emission within the same framework. 

The upper limits corresponding to the three annihilation channels $\bar{b} b, \mu^+ \mu^-, \tau^+ \tau^-$ are illustrated in Fig.~\ref{fig:IPul} (right), 
for a fixed choice of Psh+11 GMF and PDDE propagation setup.
The limits reach approximately the same value at about $500$~GeV, where the synchrotron emission spectrum from DM annihilations at $30$~GHz has similar values for all channels. 
For all channels, 
the synchrotron polarization data provide constraints at least a factor of five better than the intensity. 
For different MF (left panel) this is valid for $m_\mathrm{DM}>40$~GeV.
At tens of GeV and for $\mu^+ \mu^-$ annihilations, we exclude $\langle \sigma v\rangle$ larger than about $10^{-25}$cm$^3$s$^{-1}$. This is compared to the thermal relic cross section \cite{Steigman:2012nb} shown as a dotted line.
For the $\mu^+ \mu^-$ channel our upper limits using \planck polarization are competitive with \planck CMB constraints~\cite{Planck:2018vyg} (dot-dashed lines) between about 50~GeV and 100~GeV.
We interpret the stronger DM constraints from polarization as coming from two effects. 
First,  the astrophysical backgrounds are lower in polarization rather than in the intensity \cite{supplemental}. 
Secondly, the intensity and polarization maps have significantly different morphologies. In particular, as can be seen in Fig.~\ref{fig:planck} the polarization map presents filaments, or arms, extending many degrees in the sky. This leaves inter-arms regions with very low background very close to the Galactic center, where the DM signal peaks.
Instead, the background for the intensity has a more uniform structure towards the inner Galaxy.

While these limits on WIMPs are overall weaker than some other constraints available in the literature~\cite{Fermi-LAT:2016uux,Leane:2018kjk,Calore:2018sdx,Regis:2021glv,Cuoco:2016eej,Kahlhoefer:2021sha,2022arXiv220203076C}, the conservative analysis presented in this letter is the first step towards a more detailed assessment of the constraining power of polarization data when also the astrophysical background will be included. 
 
Further systematic uncertainties related to the choice of the propagation setup or the DM radial profile are discussed  in \cite{supplemental}.

\textbf{\textit{Conclusions.}}
This paper presents a new method to constrain DM properties using for the first time as observable the map of CMB foreground polarization. We have derived new, conservative (i.e., removing only the CMB) DM constraints using \planck synchrotron microwave polarization sky maps.
We obtain competitive bounds on the WIMP annihilation cross-section, while we find that polarization maps provide DM limits up to one order of magnitude stronger than the ones coming from intensity maps.
Our method could be generalized to other types of particles with electromagnetic annihilation or decay products. The bounds could be straightened by a proper removal of astrophysical foregrounds on top of the CMB background, by a more accurate modeling of the GMF and of the DM density profile, and finally by more sensitive  full-sky observations of the polarized millimeter sky (which should be delivered by the LiteBird satellite \cite{LiteBIRD:2020khw}).

\acknowledgments
\textbf{\textit{Acknowledgments.}}
We thank Michael Kr\"amer for insightful discussion in the initial stages of this work 
and \galprop developers for useful conversation.
We also thank Fabio Finelli and Andrea Zacchei for providing further insight on the \planck LFI maps.
The work of A.C. is supported by: ``Departments of Excellence 2018-2022'' grant awarded by the Italian Ministry of Education, University and Research (MIUR) L. 232/2016; Research grant ``The Dark Universe: A Synergic Multimessenger Approach'' No. 2017X7X85K, PRIN 2017, funded by MIUR; Research grant TAsP (Theoretical Astroparticle Physics) funded by INFN.
Simulations  were performed with computing resources granted by RWTH Aachen University.

\bibliography{biblio}


\medskip
\onecolumngrid

\setcounter{page}{1}
\renewcommand{\thepage}{S\arabic{page}} 
\renewcommand{\thetable}{S\Roman{table}}  
\renewcommand{\thefigure}{S\arabic{figure}}
\renewcommand{\theequation}{S\arabic{equation}}
\setcounter{section}{1}
\setcounter{figure}{0}
\setcounter{table}{0}
\setcounter{page}{1}
\setcounter{equation}{0}

\begin{center}
\vspace{0.05in}
{ \large {\it  Supplemental Material: 
}\\
\vspace{0.05in} Dark Matter constraints from Planck observations  of the Galactic polarized synchrotron emission}\\ 
\vspace{0.05in}
{Silvia Manconi, Alessandro Cuoco, Julien Lesgourges}
\end{center}

\begin{figure*}[b]
	\centering
	\includegraphics[width = 0.49\textwidth]{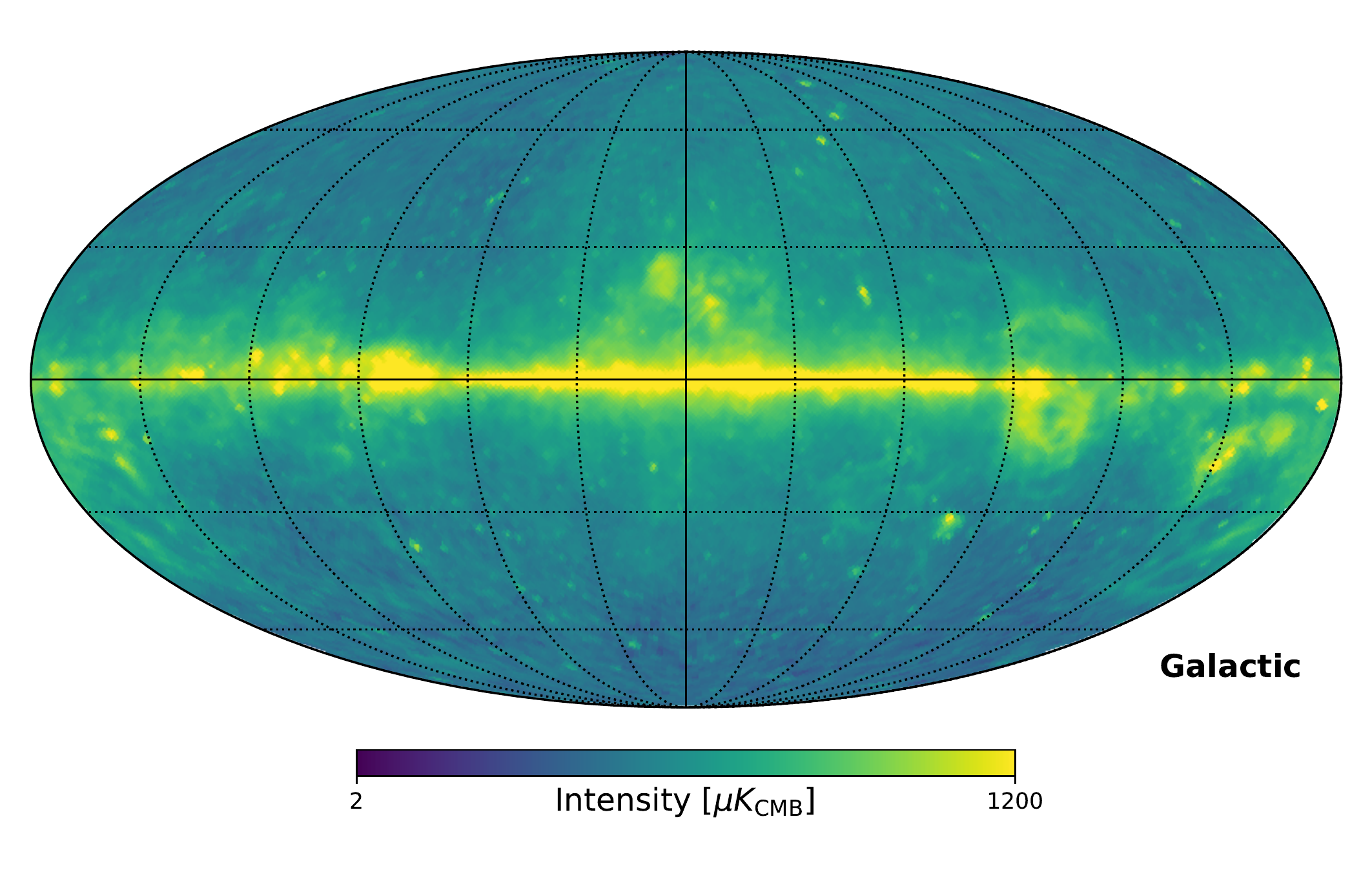}
		\includegraphics[width = 0.49\textwidth]{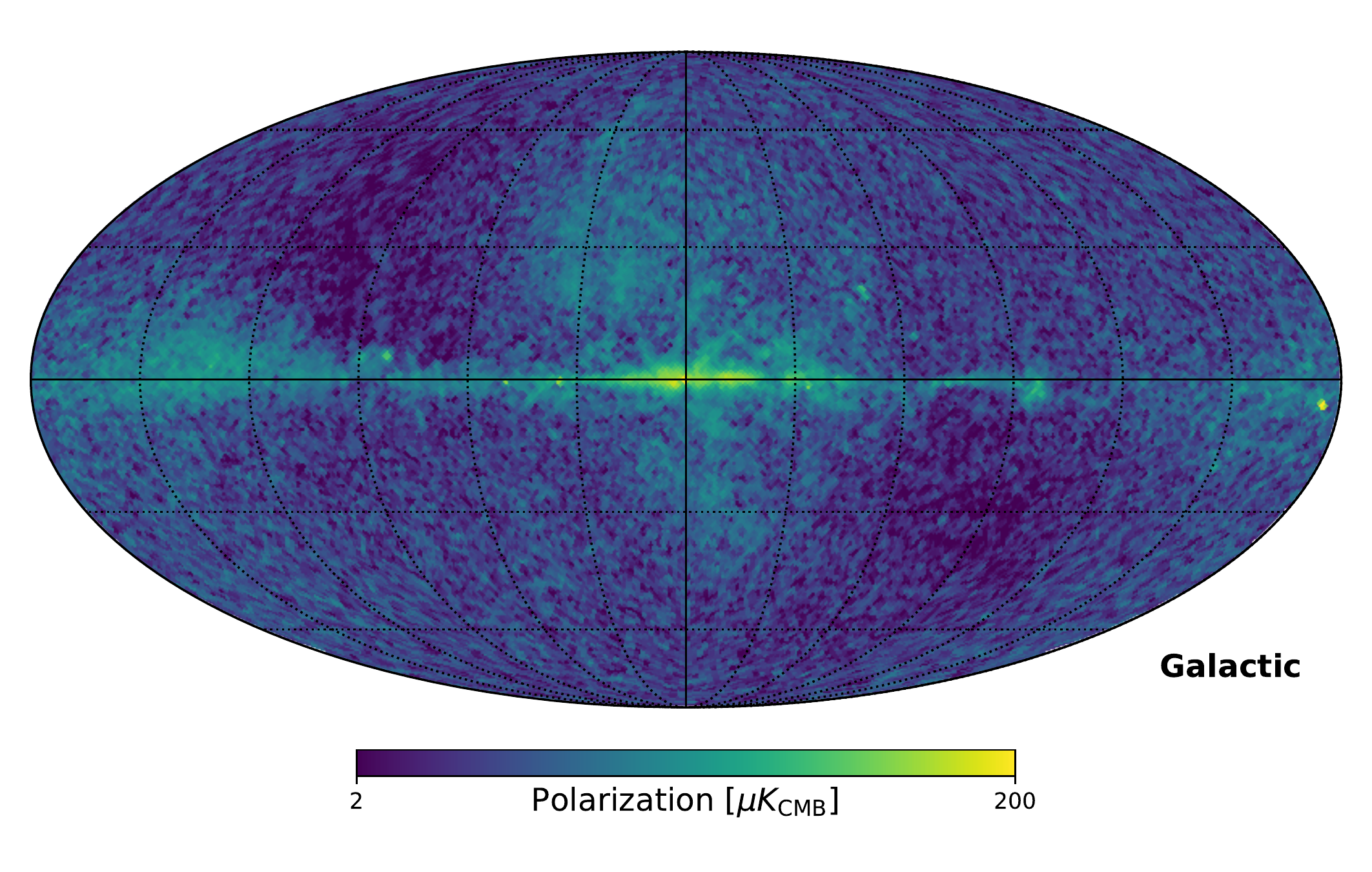}\\
				\includegraphics[width = 0.49\textwidth]{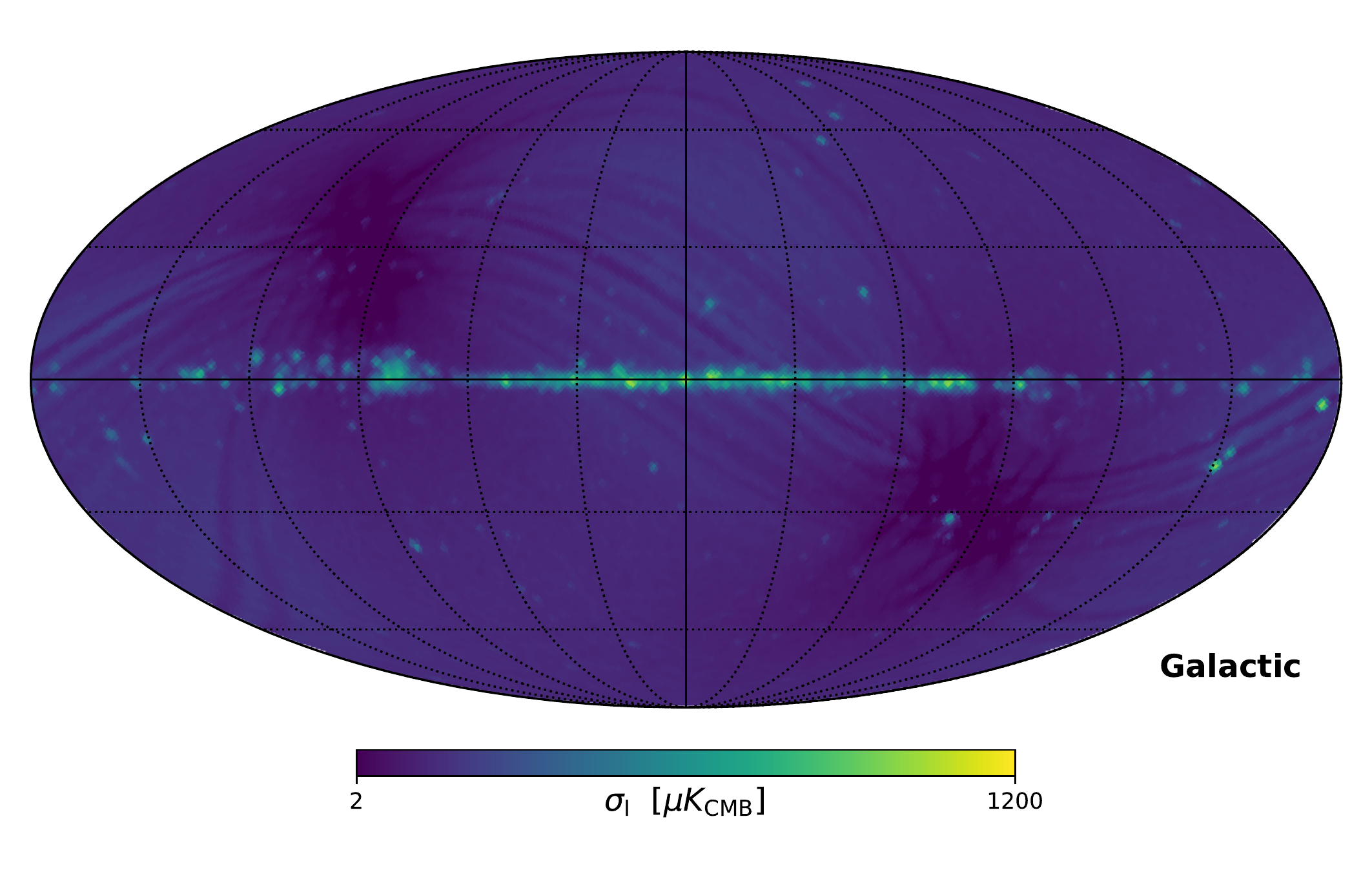}
					\includegraphics[width = 0.49\textwidth]{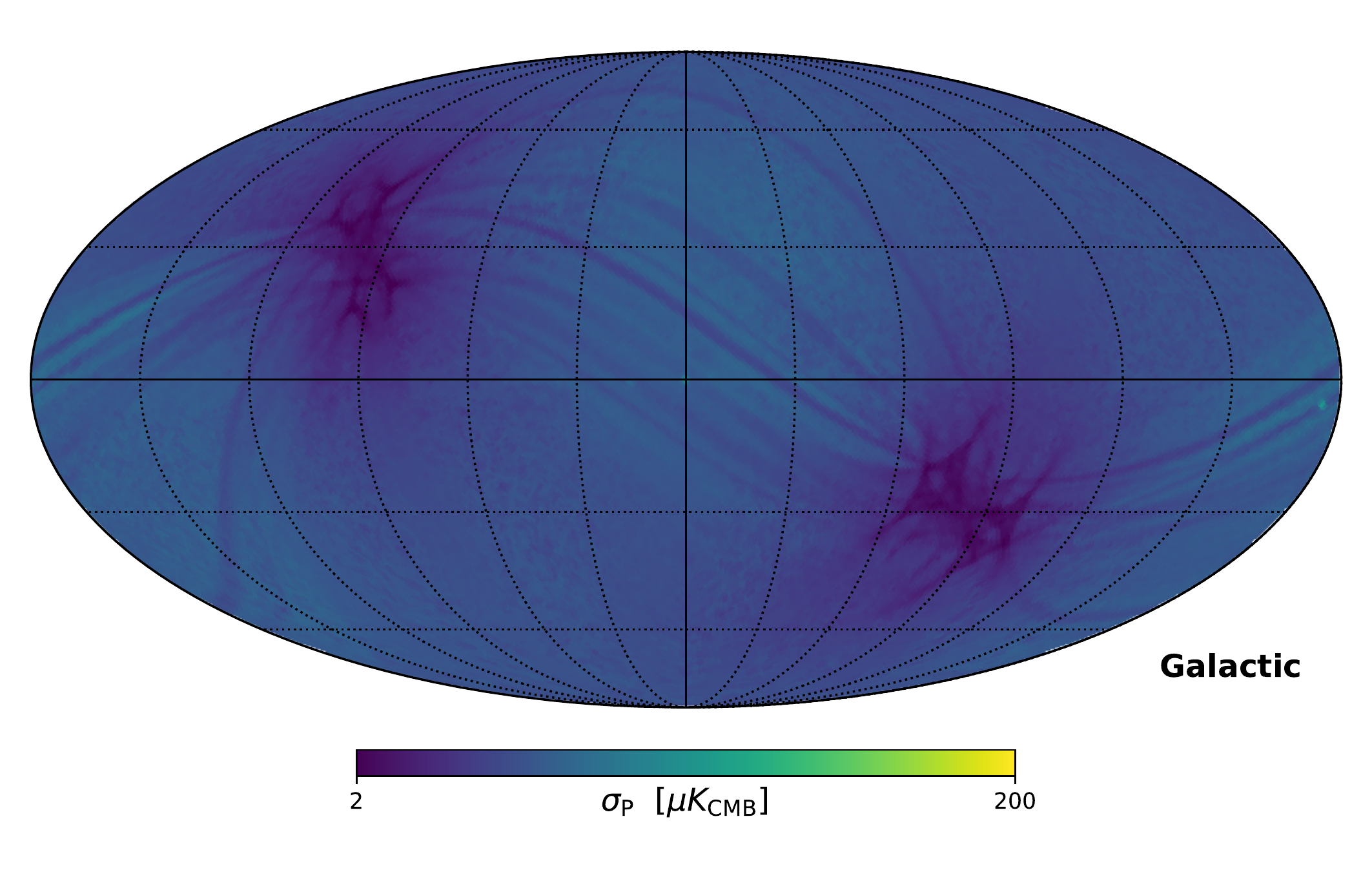}
	\caption{\textbf{Same as Fig.~\ref{fig:planck} but for \planck 44 GHz data and error estimate ($\sigma$)}.
	} 
	\label{fig:planck44}
\end{figure*}

\begin{figure*}[t]
	\centering
	\includegraphics[width = 0.49\textwidth]{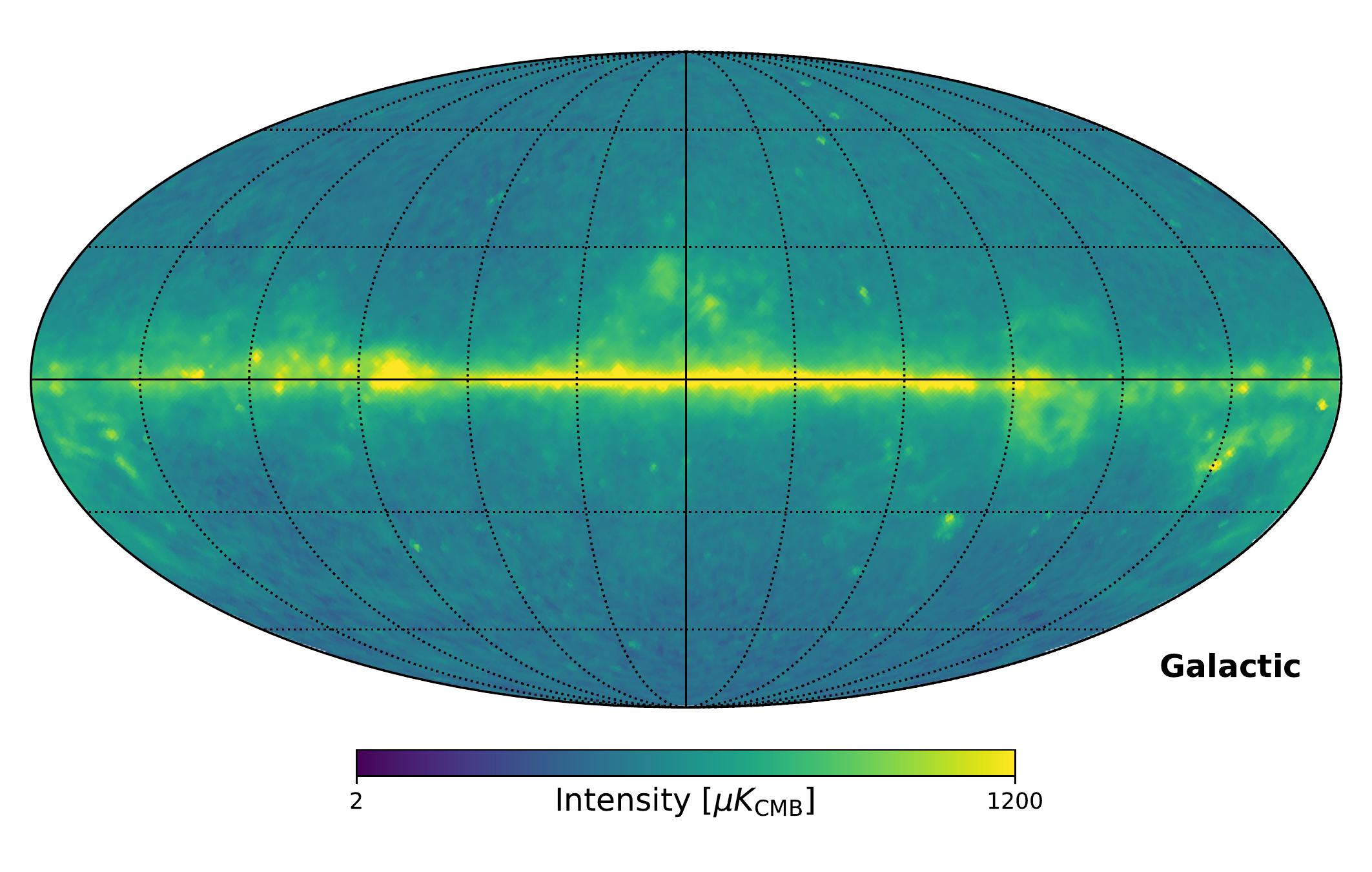}
		\includegraphics[width = 0.49\textwidth]{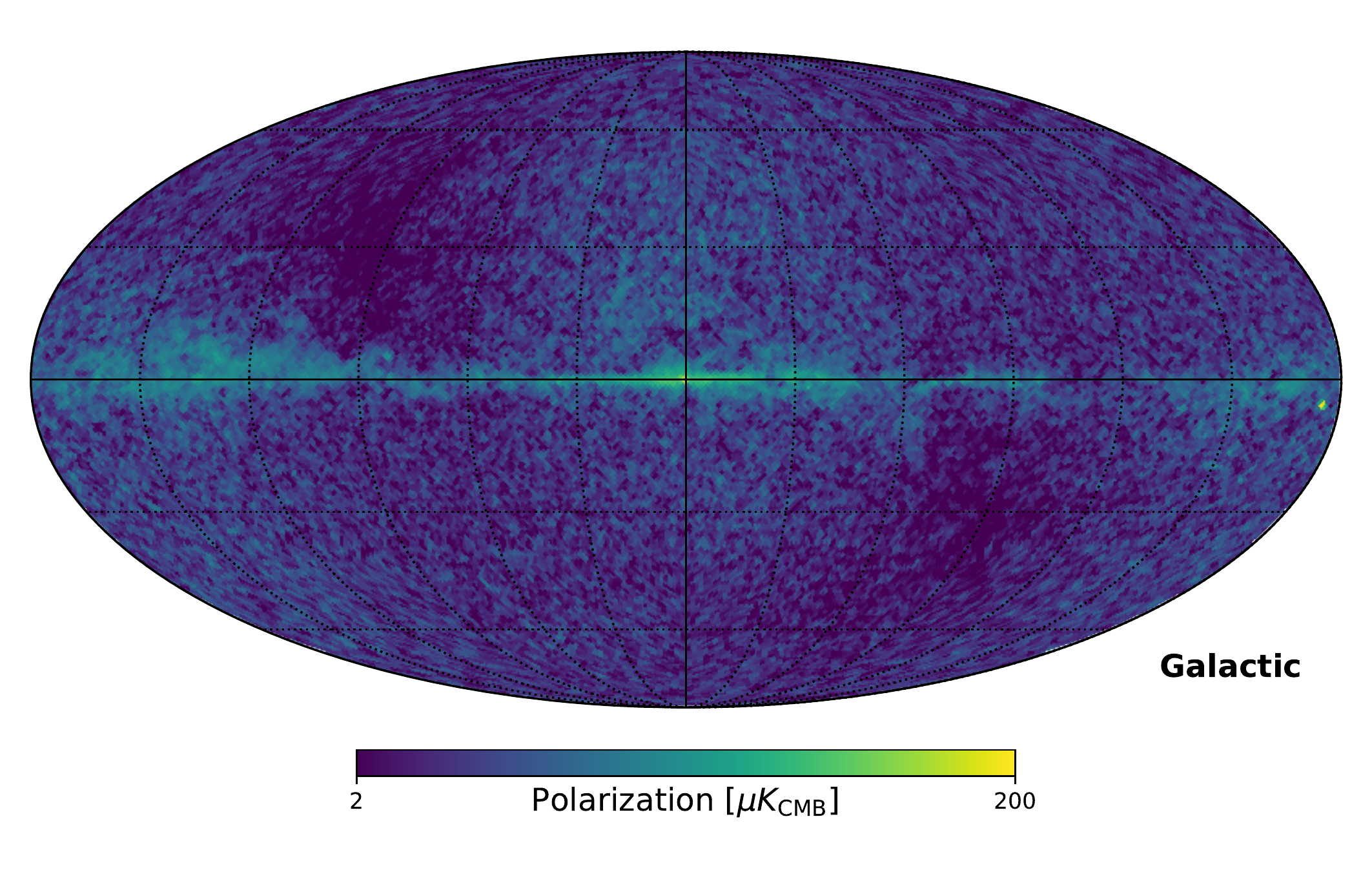}\\
				\includegraphics[width = 0.49\textwidth]{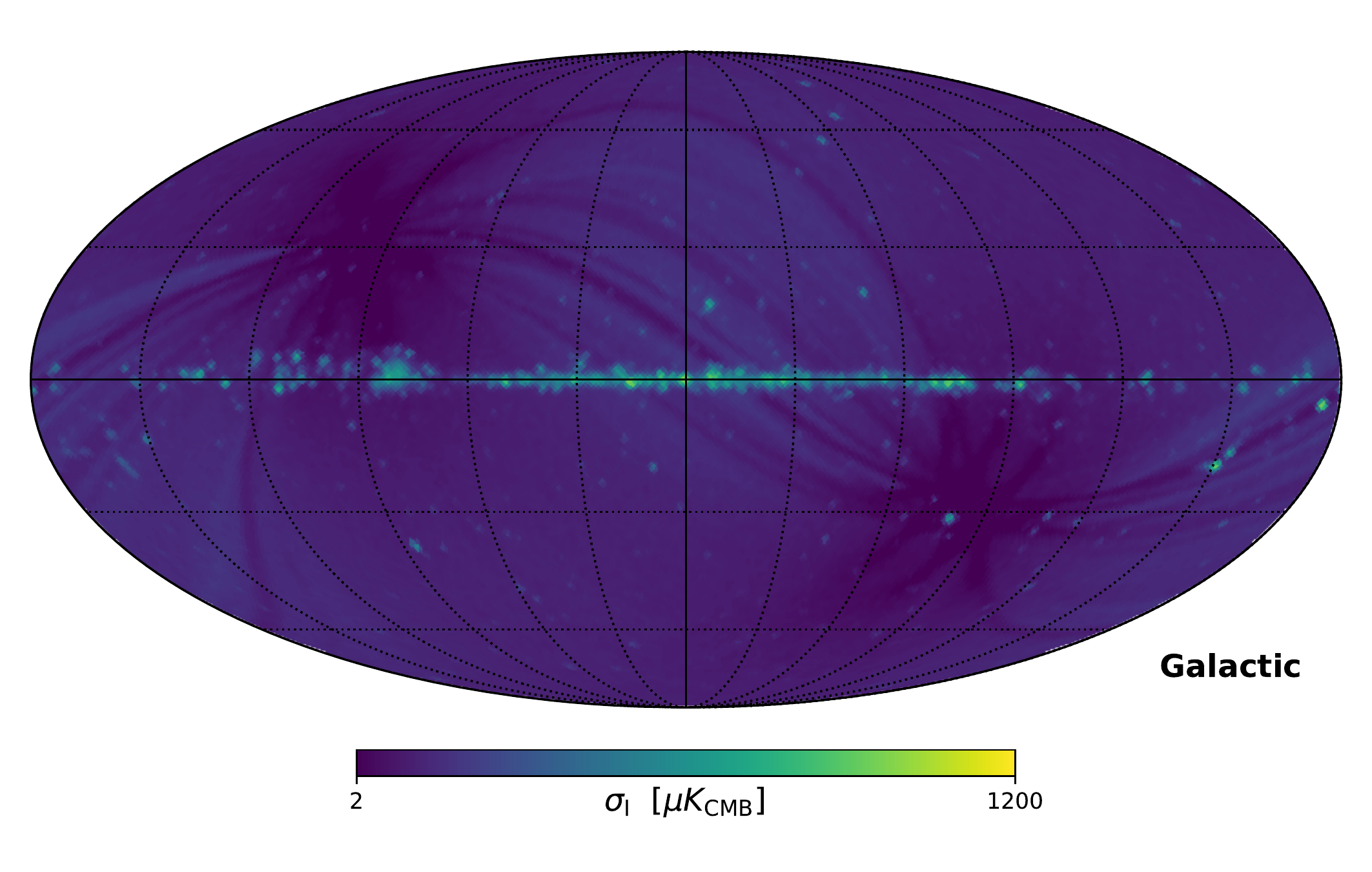}
					\includegraphics[width = 0.49\textwidth]{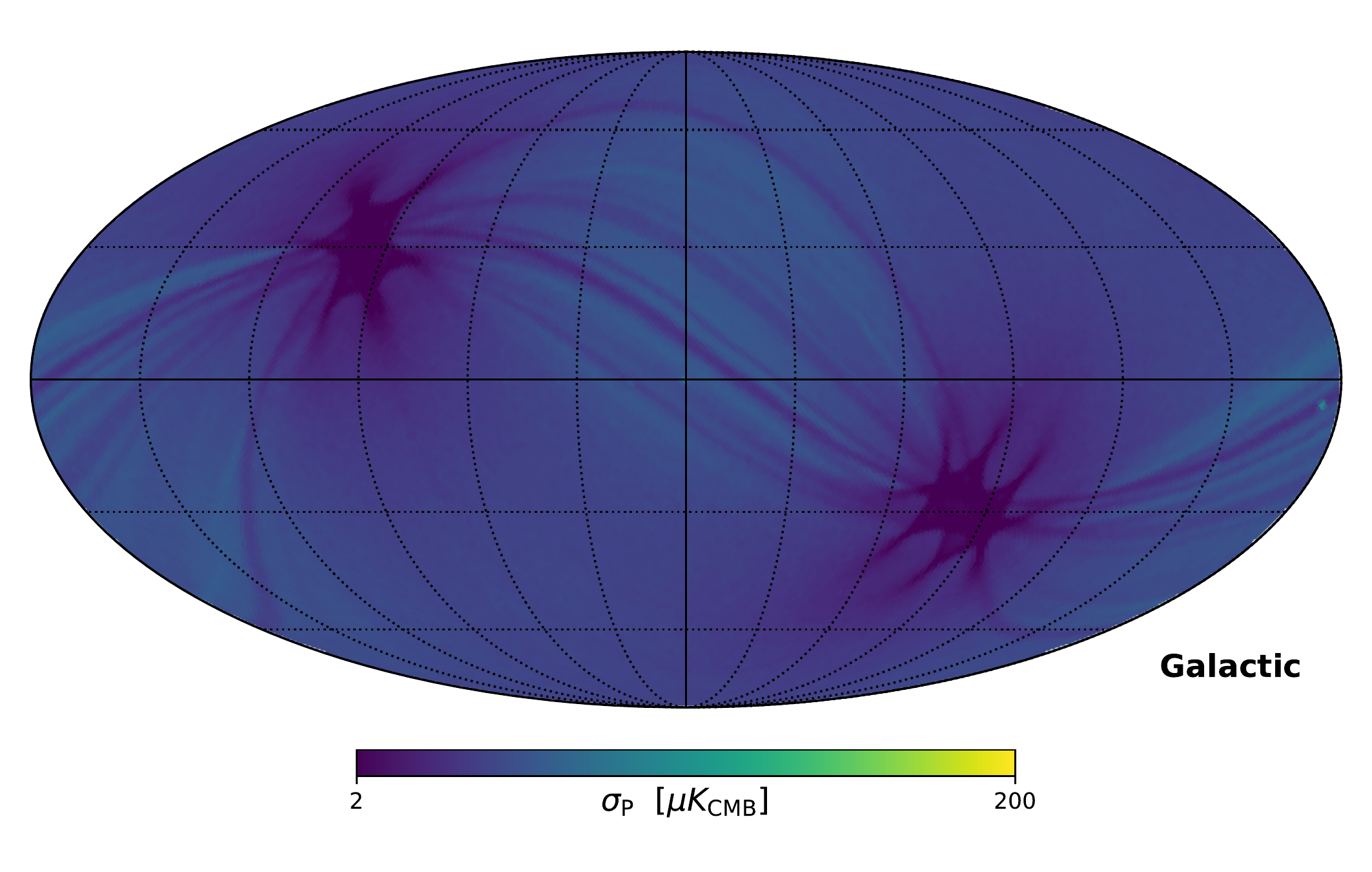}
	\caption{\textbf{Same as Fig.~\ref{fig:planck} but for \planck 70 GHz data and error estimate ($\sigma$)}.
	} 
	\label{fig:planck70}
\end{figure*}

\section{More on microwave maps}\label{sec:mwplanck}
In Fig.~\ref{fig:planck44}-\ref{fig:planck70} we illustrate, similarly to Fig.~\ref{fig:planck} in the main text, the intensity and polarization amplitude as measured by \planck and the error estimates for the other two LFI frequencies of $44$~GHz and $70$~GHz. 

The temperature $T_{I, P,  \rm CMB}(\nu)$ in units of $K_{\rm CMB}$ and  the brightness temperature $T_{I, P}(\nu)$ in units of $K_{\rm RJ}$ are related as: 

\begin{equation}\label{eq:Tcmb}
 T_{I, P, \rm CMB}(\nu)= T_{I, P}(\nu) \frac{1}{\mathcal{C}_c(\alpha)} \frac{(e^{x_c}-1)^2}{x_c^2 e^{x_c}}
\end{equation}
where  $x_c=\frac{h \nu}{k_B T_{\rm CMB}}$, $\nu$ is the central frequency of the  considered channel (e.g. 30 GHz), $T_{\rm CMB}=2.7255$K, $k_B$ is the Boltzmann constant and $C_c(\alpha)$ is a color correction factor, equal to  $\mathcal{C}_\mathrm{30GHz}(-1)=0.969$ for synchrotron radiation and the $30$~GHz channel, see  Ref.~\cite{2016A&A...596A.103P} for more details. 

Before comparing the DM predicted polarization $P$ with the observed $P$ we need to build the experimental $P$ map and its error map from the $Q$ and $U$ maps.
We here provide additional details on the construction of  
these maps.  
As anticipated in the main text, after the first \textit{smoothing} step, we proceed with the 
 \textit{error estimation.}
Using the (unsmoothed) $I,Q,U$ maps at the native NSide resolution, for each pixel we consider neighboring pixels up to 0.5 degrees and we use them to compute the variance and the $\sigma$ in that pixel. 
The resulting $I,Q,U$ error maps are then  smoothed  with a Gaussian beam of FWHM of $1$~deg to remove some residual small scale noise.
The error map for $P$ is derived at this point from the error maps on $Q,U$ using error propagation:
\begin{equation}
    \sigma_{\rm P} =\sqrt{\frac{U^2}{P^2} \sigma_{\rm U}^2 + \frac{Q^2}{P^2} \sigma_{\rm Q}^2 }\,.
\end{equation}
Comparing the error maps with the $I,P$ maps in the upper panel of Fig.~\ref{fig:planck}, we note that, as expected, the $P$ map has an overall larger relative error. We provide the resulting sky maps for the error for $44$~GHz and $70$~GHz in Fig~\ref{fig:planck44} and Fig~\ref{fig:planck70}. 
We note that this procedure returns the total variance, which is the sum of the intrinsic variance of the map (which is anisotropic due to the morphology of the Galactic backgrounds) plus the noise. Nonetheless, at the small scales where we perform the calculation the map is dominated by noise. Thus, the obtained variance is a good estimate of the noise, i.e., of the error. 

Finally, in order to compare \planck data to theoretical predictions for the DM annihilation signal, we proceed by \textit{degrading} the $I,P$ maps to low resolution using the \texttt{healpy.ud\_grade} routine. This procedure averages the high-resolution pixels in each lower resolution ones.  We note that this might be problematic for non-scalar quantities such as $P$, since the used \texttt{healpy} routine does not include parallel transport. However, this causes only small uncertainties close to the coordinates poles, and has no impact for the current analysis.  
When degrading the error maps, we must take into account the fact that they represent an error {\it per pixel}. Thus they need to be rescaled if the pixel size is changed.
Assuming we are not in a systematic-limited regime, the error will scale with the number of observations, i.e. the number of pixels. Since at each degrading step four pixels are grouped together, going from NSide 1024 to NSide 512 a rescaling factor of $1/\sqrt{4}=1/2$ must be applied, i.e., the signal-to-noise increases for larger pixels.
For a generic lower resolution $l$, the rescaling factor reads $\sigma_{\rm P}^{l}=\sigma_{\rm P}/\sqrt{N_{\rm pix}^{1024}/N_{\rm pix}^{l}}$. 

We have tested that using the CMB-subtracted maps 
obtained using other methods (COMMANDER, SEVEM, or SMICA) instead of NILC (see main text) would change our results only by few percent.
Subtracting the CMB avoids the need to deal with negative values in intensity maps arising from regions in which the CMB has cold temperature fluctuations not compensated by astrophysical backgrounds or by noise.
This choice has only a minor impact on DM limits from synchrotron polarization, since the CMB signal is anyway subdominant in polarization maps at the low frequencies we are interested in.
We note that this is a difference with respect to some previous works computing conservative DM limits (using total intensity only, see e.g. Refs.~\cite{Egorov:2015eta,Cirelli:2016mrc}) which typically use the native maps without subtracting the CMB contribution. 

Since some negative pixels are left after the preprocessing of the intensity map at large latitudes in the southern hemisphere, we cut latitudes $|b|>30$~deg. We checked that for polarization this is not changing the results, since the most constraining pixel is always lying at $|b|<30$~deg towards the Galactic center, where the DM signal is larger (see Fig.~\ref{fig:mapPdm}).

\section{Synchrotron emission from Galactic dark matter}\label{sec:model}
Galactic synchrotron is among the main diffuse emissions observed by \planck LFI, both in the total intensity and in polarization. At frequencies below about $50$~GHz, the total intensity also contains significant contributions from the free-free emission coming from bremsstrahlung in electron-ion collisions and from spinning dust. 
At frequencies above about $100$~GHz, the thermal dust emission is expected to dominate both the total intensity and the polarization signal \cite{Planck:2018nkj}. 
In what follows we describe how we compute the potential synchrotron emission produced by Galactic DM annihilation, leaving an extended analysis of the DM signal together with astrophysical backgrounds for future work.

The DM synchrotron emission for each Stokes parameter  $I,P$ (in units of erg cm$^{-2}$/s/Hz/sr ) can be written as:
\begin{equation}\label{eq:Jmain}
 \mathcal{J}_{I, P}(\nu, b, l)= \frac{1}{4\pi} \int_{\rm los} ds \int dE \,  \mathcal{N}_{\rm e^\pm} (E, \mathbf{x}) \mathcal{P}_{I, P}(\nu, E)\,.
\end{equation}
The second integral in the $e^\pm$ energy $E$ depicts the synchrotron emissivity for a cell located at position $\mathbf{x}$ along the line of sight. This is obtained by convolving the $e^\pm$ number density  $\mathcal{N}_{\rm e^\pm} (E, \mathbf{x})$  at Galactic position $\mathbf{x}$ (which depends on DM properties and $e^\pm$ propagation) with the synchrotron emission power $\mathcal{P}_{I, P}(\nu, E)$  emitted at frequency $\nu$ by relativistic $e^\pm$ with energy $E$ (which depends on the GMF properties).
This emissivity is then integrated spatially  over the line of sight (los) distance $s$, individuated by the Galactic coordinates $b, l$.
The synchrotron intensity and polarization fluxes in Eq.~(\ref{eq:Jmain}) can be expressed as \textit{brightness temperature}, see Eq.~(\ref{eq:Tb}). 

\medskip 
As anticipated in the main text, we use \texttt{GALPROP} version v54r2766  as adapted  in Ref.~\cite{Egorov:2015eta} to compute all-sky  synchrotron signal maps from DM annihilations defined by Eq.~(\ref{eq:Jmain}). 
The computation of the total synchrotron intensity and polarization amplitude is based on the  \texttt{GALPROP} developments described in Ref.~\cite{strong2011,Orlando:2013ysa}, and includes free-free absorption, which  we recall is however expected to be subdominant at \planck frequencies. 
In what follows we complement the main text by detailing the assumptions used to compute each term in Eq.~(\ref{eq:Jmain}). 

The DM modeling is based on the one discussed in Ref.~\cite{Egorov:2015eta}, with modifications described in
what follows. We then detail the \texttt{GALPROP} configuration used to solve the transport of $e^\pm$ from DM in the Galaxy and the propagation models explored. Finally, we summarize the GMF models employed in this study.

\subsection{Dark matter modeling}\label{subs:dm}
The distribution of DM in our Galaxy is still poorly constrained, especially within the Solar circle \cite{Benito:2019ngh}.
Here, we assume the DM density profile in the Galaxy entering in Eq.~(\ref{eq:qdm}) to be spherical symmetric, and, to gauge the uncertainties associated with it, we consider  different radial profiles. 
As a benchmark, we consider the DM radial distribution $\rho_{\rm DM}(r)$ in the Galaxy at distance $r$  from the halo center to be described by a standard NFW density profile \cite{Navarro:1995iw,Navarro:2008kc}: 
\begin{equation}
    \rho_{\rm DM}(r) = \frac{\rho_s}{ \left( \frac{r}{r_s} \right)^\gamma  \left(  1+ \left( \frac{r}{r_s}\right)^{\alpha} \right)^{(\beta-\gamma)/\alpha} }~,
\end{equation}
where $\rho_s$ and $r_s$ are the density and the scale radius. The parameters $\alpha, \beta, \gamma$ determine the shape of the profile, and are fixed as $\alpha=1, \beta=3, \gamma=1$.
We fix $r_S=23$~kpc and the local DM density at the solar position to be $\rho_{\rm DM}(r_\odot =8.5~\rm{kpc})=0.4$~GeV/cm$^3$ \cite{deSalas:2020hbh}. 
To estimate the uncertainties related to the DM radial distribution, particularly relevant for the innermost part, we also consider two additional cases. 
Dark matter profiles steeper than $\gamma=1$, such as the so-called generalized NFW (gNFW) profiles with $\gamma>1 $ can accomodate baryonic effects in  simulations of cold DM, and are currently suggested by most analysis interpreting the GeV excess in the inner Galaxy in terms of DM annihilations \cite{Murgia:2020dzu}. 
We thus consider a gNFW with $\gamma = 1.25$ and $r_S=27.2$~kpc, see also Ref.\cite{Egorov:2015eta}. 
Finally, a cored Burkert profile \cite{Burkert:1995yz} is considered, with $r_s=12.67$~kpc.
All the profiles are normalized to the same local DM density 
$\rho_{\rm DM}(r_\odot =8.5~\rm{kpc})=0.4$~GeV/cm$^3$ as illustrated in the left panel of Fig.~\ref{fig:DMprofile}.
We see that the cored, Burkert profile predicts a much smaller DM density in the innermost part of our Galaxy, while the NFW and the gNFW have a much steeper density profile. This has a significant impact on the constraints on the DM synchrotron signal, as demonstrated in the right panel of Fig.~\ref{fig:DMprofile}, and further discussed below.

\subsection{Cosmic-ray propagation}\label{subs:galprop}
The \texttt{GALPROP} setup we use to numerically solve the transport equation and compute the number density of $e^\pm$ from DM annihilations at each position in the Galaxy includes spatial  diffusion with an isotropic and spatially-independent diffusion coefficient, diffusive reacceleration in the interstellar medium, convection from the Galactic wind,  energy losses via ionization, Coulomb losses, bremsstrahlung,  synchrotron radiation and inverse Compton scattering on the interstellar radiation fields.
We refer to Refs.\footnote{https://sourceforge.net/projects/galprop/}-\cite{Egorov:2015eta} for the description and the implementation in  \texttt{GALPROP} of these different processes. 
We recall that this specific public version of \texttt{GALPROP} was adapted  in Ref.~\cite{Egorov:2015eta} to introduce a DM source of CRs following Eq.~(\ref{eq:qdm}).

In order to solve the propagation equation, the parameters of the diffusion model have to be specified.
Many recent works used the wealth of high-precision AMS-02 CR data \cite{AGUILAR20211} to constrain the available parameter space, see e.g. \cite{2018MNRAS.475.2724O,2019PhRvD..99d3007O,Korsmeier:2021brc}.
We summarize in Tab.~\ref{tab::diff} the parameters of the three models employed to bracket the uncertainties related  the propagation (PDDE, DRE, BASE, see main text).
Note that the diffusion coefficient is defined as  $D_{\rm xx}=10^{28}\beta D_0(R/D_R)^{\delta}$ cm$^2$s$^{-1}$, and is normalized at $D_R=40$~GV for PDDE and DRE models and $D_R=4$~GV for BASE model.
The parameters $D_{\rm bf}, \delta_1, \delta_2$ describe the break in the diffusion coefficient, $V_{\rm Alf}$ is the Alfven velocity for the reacceleration term, $v_{\rm c}$ is the convection velocity and $z$ the half-width of diffusion halo of the Galaxy.
We refer to the original publications \cite{2018MNRAS.475.2724O,2019PhRvD..99d3007O,Korsmeier:2021brc} for a detailed description of these propagation models and their compatibility with CR and multiwavelength data. 

\begin{table}[t]
\caption{Diffusion models. Notes: the diffusion coefficient is defined as  $D_{\rm xx}=10^{28}\beta D_0(R/D_R)^{\delta}$ cm$^2$s$^{-1}$, and is normalized at $D_R=40$~GV for PDDE and DRE models and $D_R=4$~GV for BASE model.}
\centering
\begin{tabular}{ l @{\hspace{10px}} c @{\hspace{10px}} c @{\hspace{10px}} c  } \hline\hline
Model parameters  &   \textbf{PDDE}          &   \textbf{DRE}  &  \textbf{BASE}     \\ \hline
$D_0$ [cm$^2$s$^{-1}$]   &   $12.3$  &  $14.6$&  $5.05$ \\
$D_{\rm br}$ [GV]   &   $4.8$  &  - & $4.0$ \\
$\delta_1$  &   $-0.641$  &  $0.327$&  $-0.98$ \\
$\delta_2$  &   $0.578$  &  $0.323$&  $0.49$ \\
$V_{\rm Alf}$ [km s$^{-1}$] &  -  &  $42.2$&  - \\ 
$v_{\rm c}$ [km s$^{-1}$] &  -  &  - &  $3.34$ \\
$z$ [kpc] &  $4$  &  $4$&  $4$ \\
\hline\hline
\end{tabular}
\label{tab::diff}
\end{table}

\subsection{Galactic magnetic field models}\label{subs:Bfield}
We here briefly recall some basic concepts and list the specific models we explored, while we refer the reader to Ref.~\cite{jaffe2019} (and references therein) for a comprehensive review.
The intensity and spatial structure of the GMF are still uncertain, and currently constrained through Faraday rotation measurements of pulsars and extragalactic source, and surveys of diffuse  synchrotron emission and polarization at radio and microwave frequencies \cite{jaffe2019}. 
The state of the art is represented by several models, which share common features such as 3D spiral structures in disks, but cover a variety of morphologies for the regular and random components, and constrained using different quantitative approaches. 
Previous works estimating the synchrotron intensity signals from DM often shaped the GMF by a double-exponential for the sake of simplicity \cite{Egorov:2015eta,Cirelli:2016mrc}. 
However, the polarized signal is ruled by the regular (and striated) component, and thus a much more refined description of its large scale structure is required.
The polarized signal from standard astrophysical sources has been studied in detail  \cite{Orlando:2013ysa, 2018MNRAS.475.2724O, 2019PhRvD..99d3007O,2016A&A...596A.103P} by using a number of state-of-the-art GMF models. 
In particular, CR propagation models and GMF parameters where fitted together to reproduce  multiwavelength data, including WMAP and \textit{Planck}.  This is particularly important, e.g., to tune the normalization of the random magnetic field component, which is degenerate with the normalization of CR leptons. 

As anticipated in the main text, we consider three models for the GMF, which are detailed in what follows.  

\begin{itemize}

 \item The Sun+10 model proposed in Refs.~\cite{Sun:2007mx,sun2010} describes the regular disk field as an axisymmetric spiral plus reversals in rings (ASS+RING model), and the halo field as a double torus. 
 The GMF implementation and parameters are taken as described and fitted to multiwavelength data in Ref.~\cite{Orlando:2013ysa} and recently updated in  Ref.~\cite{2018MNRAS.475.2724O}.  In particular, the regular fields have $B_{0,\rm disk,halo}=2.7 \mu$G. We take the striated component to be negligible or very low according to Ref.~\cite{2018MNRAS.475.2724O}. 
 
 The random component is instead modeled as a simple exponential law $B_{\rm ran}=B_{0,\rm ran} \exp((-R -R_{\odot})/R_{0,\rm ran}) \exp(-|z|/z_{0,\rm ran})$ with 
$B_{0,\rm ran}=4.9 \mu$G, $z_{0,\rm ran}=4$~kpc and $R_{0,\rm ran}=30$~kpc~\cite{Orlando:2013ysa,2018MNRAS.475.2724O}.

\item In the model proposed in Ref.~\cite{2011ApJ...738..192P} (Psh+11) the regular field in the disk is characterized  as a logarithmic bisymmetric spiral (BSS model \cite{2011ApJ...738..192P}), and the halo field as an asymmetric halo. Also this implementation is based  on Ref.~\cite{Orlando:2013ysa} and updated according to  Ref.~\cite{2018MNRAS.475.2724O}, with $B_{0,\rm disk,halo}=2.7 \mu$G.  
We note that in Ref.~\cite{2018MNRAS.475.2724O} only the parameters of the Sun+10  model have been updated. However, comparing with the earlier results in Ref.~\cite{Orlando:2013ysa}, in which both models have been fitted, the values of $B_{0,\rm disk,halo}$ and $B_{0,\rm ran}$ are overall similar between the Psh+11 and the Sun+10 models.
The random component is modeled as for Sun+10. 

\item A more sophisticated model has been presented by Jansson \& Farrar  for the regular \cite{Jansson:2012pc} and random magnetic fields \cite{Jansson:2012rt} (JF12).
The work of JF12 describes the regular field as the superposition of a disk, a toroidal halo and a X-field, plus a striated  component and a detailed model for the random field strength. 
We base our implementation in \texttt{GALPROP} on the one provided by Ref.~\cite{Fornengo:2014mna} and set all the parameters as in the original model for the regular \cite{Jansson:2012pc}, striated and random \cite{Jansson:2012rt} fields.  We note that a non-zero striated component and a larger strength of the random field are predicted by this model compared to the other two.  
\end{itemize}

\begin{figure*}[t]
	\centering
	\includegraphics[width = 0.49\textwidth]{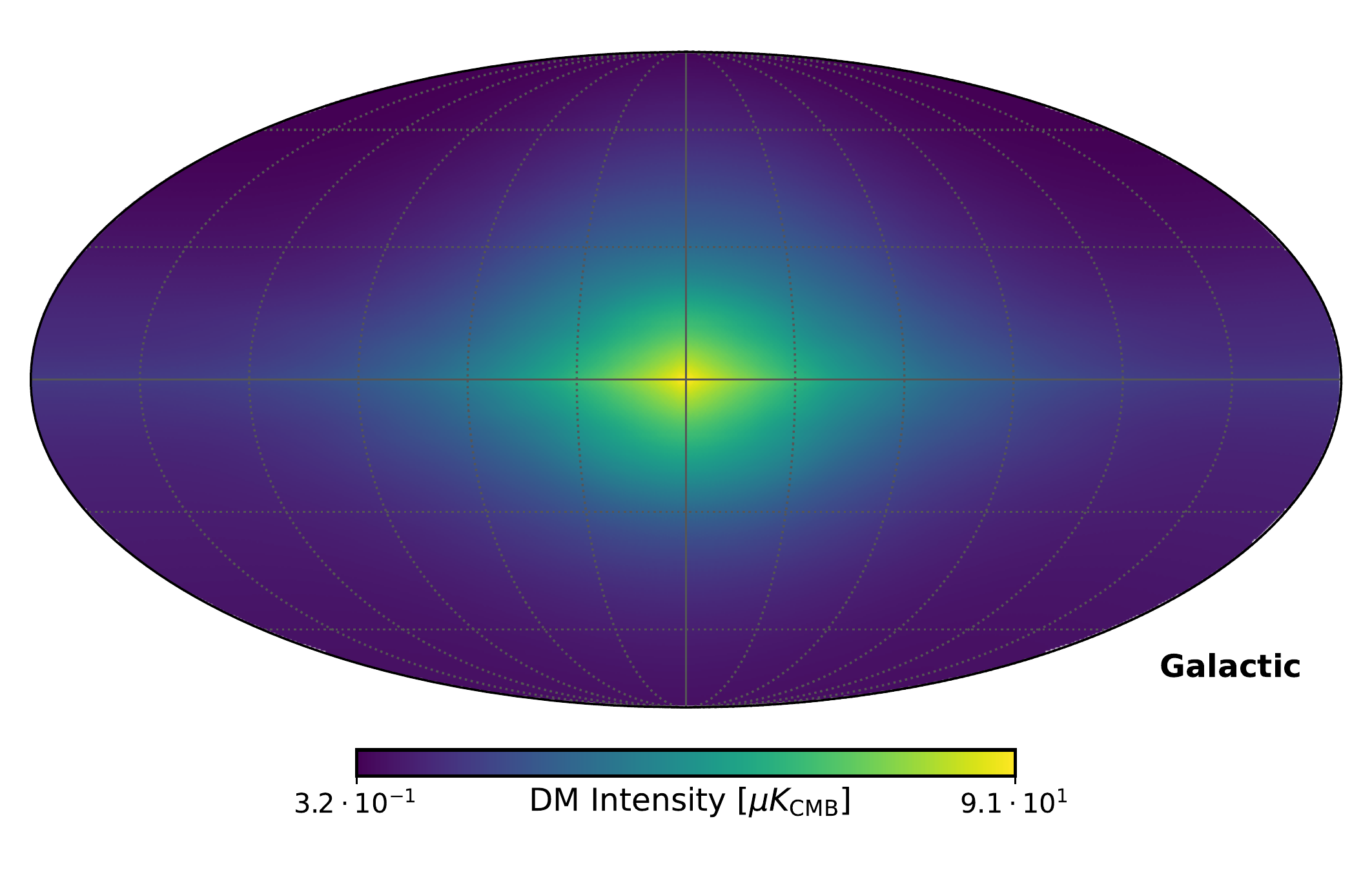}
		\includegraphics[width = 0.49\textwidth]{figures/DMmaps/PSH11/PDDE_PSH_3D_mu_50GeV_30GHz_synchrotron_P_Tb_DM_final_log_paper.pdf}
			\includegraphics[width = 0.49\textwidth]{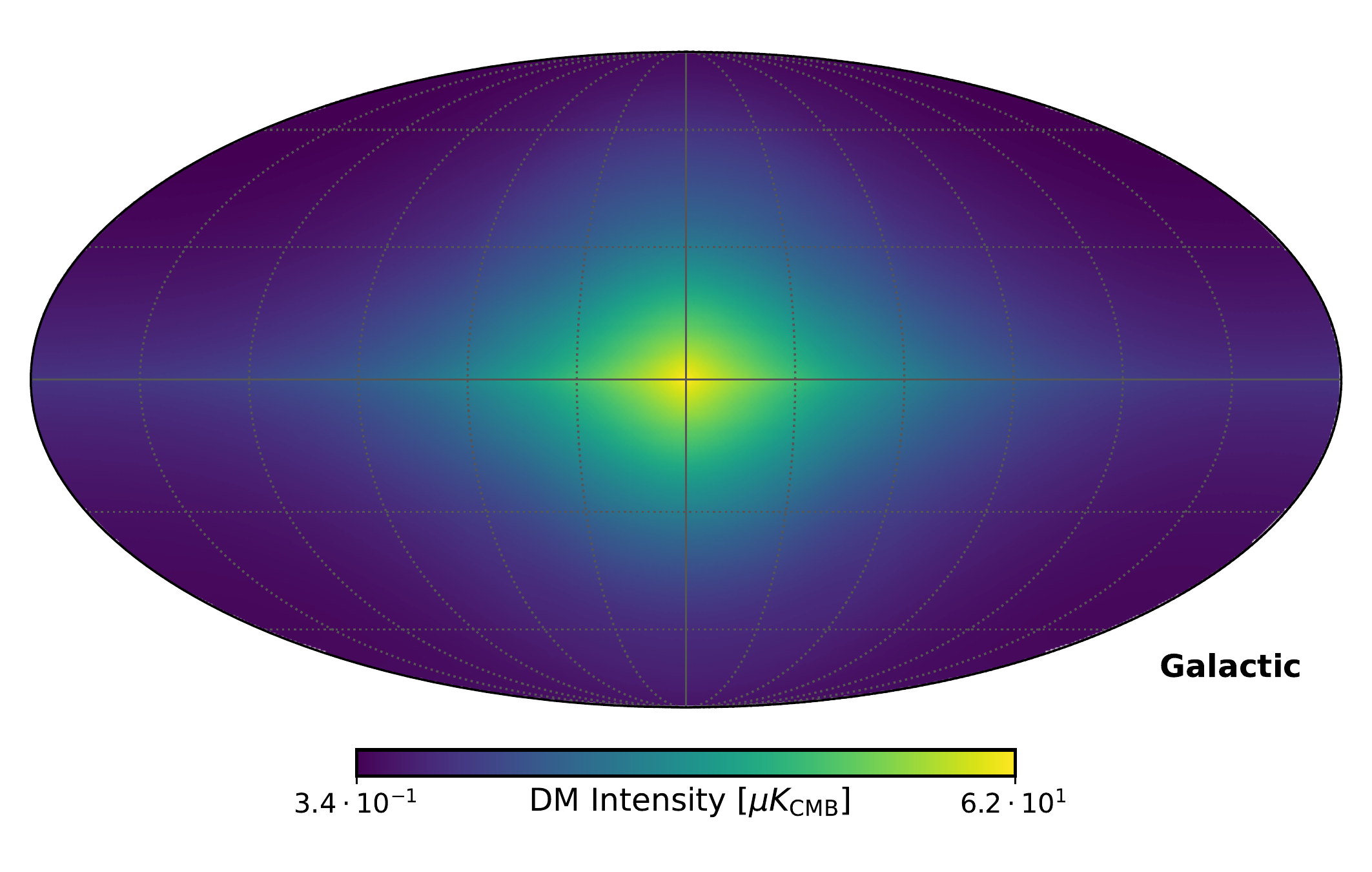}
		\includegraphics[width = 0.49\textwidth]{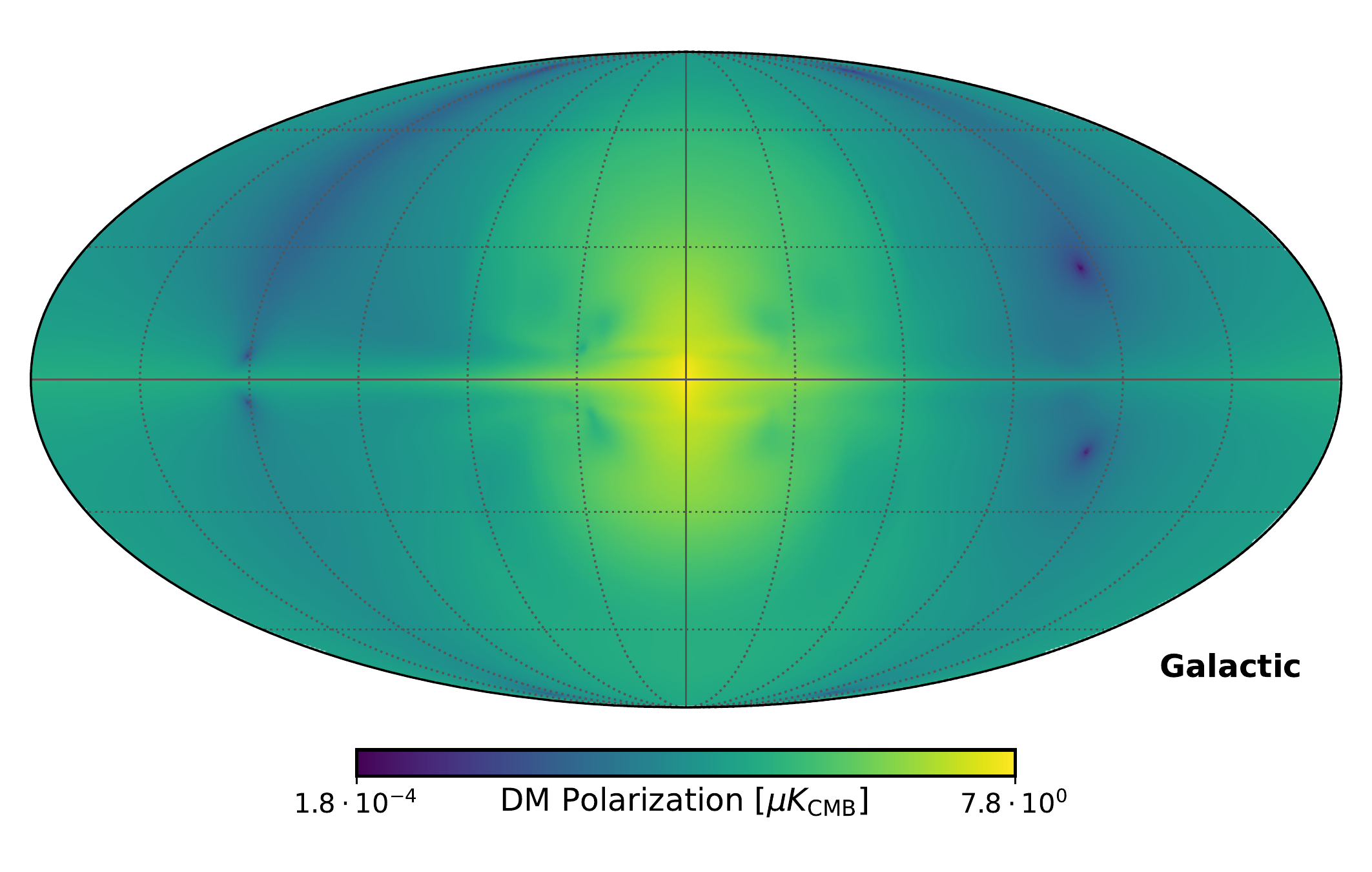}
				\includegraphics[width = 0.49\textwidth]{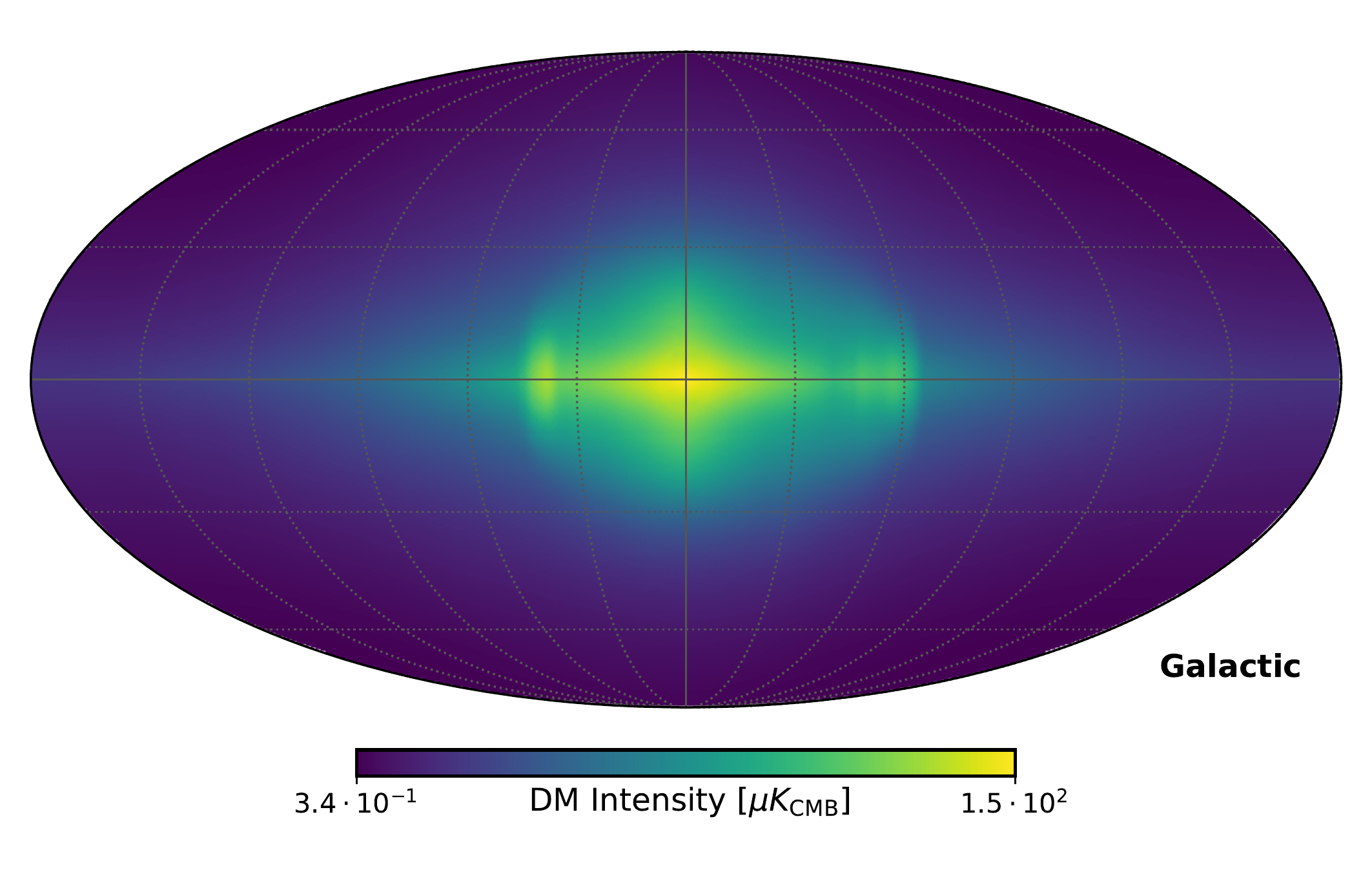}
		\includegraphics[width = 0.49\textwidth]{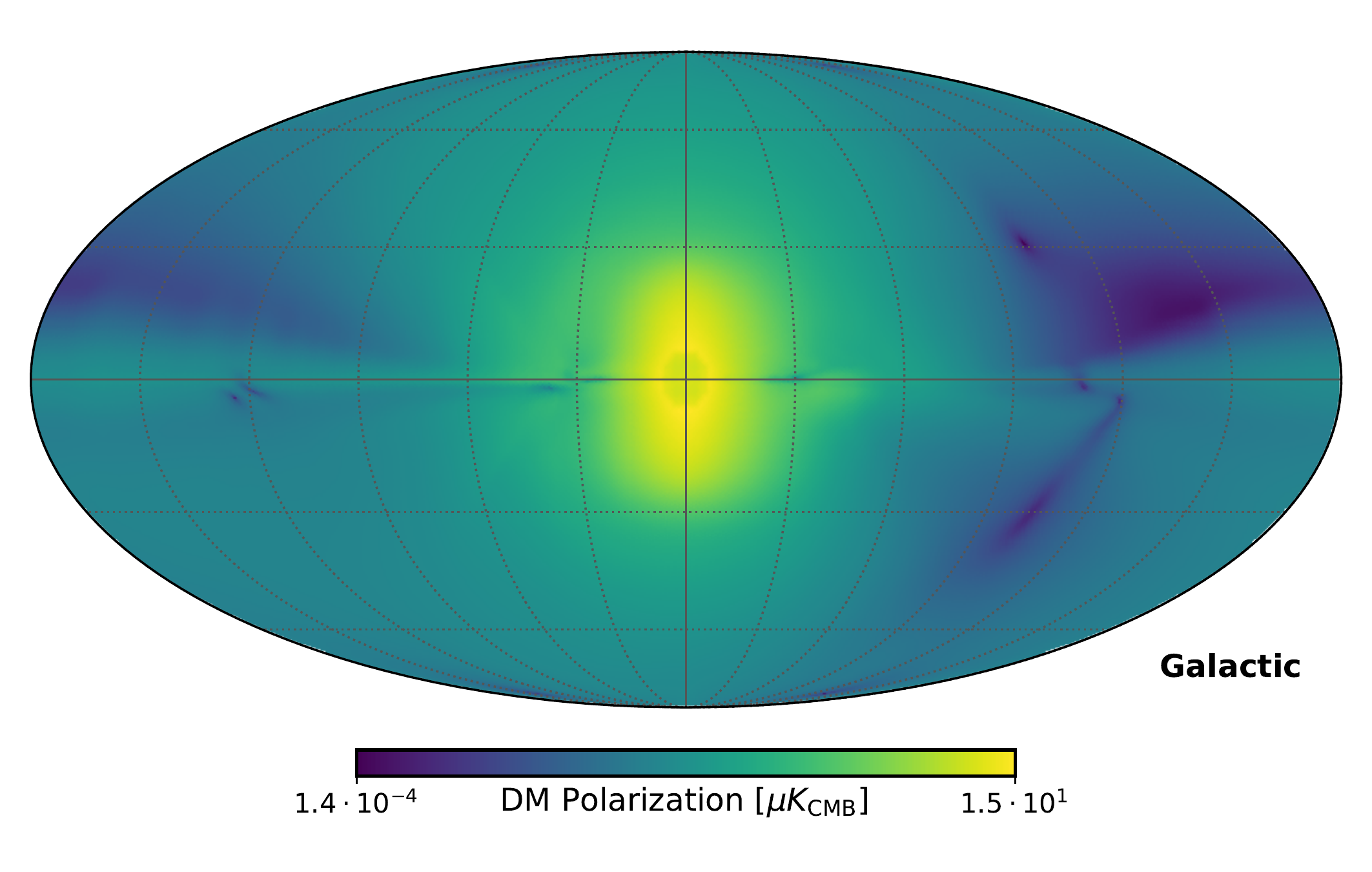}
	\caption{\textbf{Synchrotron emission from DM  at 30 GHz} as computed for DM particles of mass $m_{\rm DM}=50$~GeV annihilating into  $\mu^+ \mu^-$ pairs with a thermal averaged cross section of $\langle \sigma v\rangle = 3 \times 10^{-26}$ cm$^3$s$^{-1}$. The maps are computed with \texttt{GALPROP} using the PDDE propagation parameters \cite{2018MNRAS.475.2724O,2019PhRvD..99d3007O} and the Psh+11 
	(upper row), Sun+10 (middle) and JF12 (lower row) GMFs, and represent the Stokes intensity (left)  and polarization amplitude (right).  All maps are computed for NSide=128 and are shown in a Mollweide projection with a logarithmic color mapping.}
	\label{fig:mapIPdm}
\end{figure*}

\begin{figure*}[t]
	\centering
	\includegraphics[width = 0.49\textwidth]{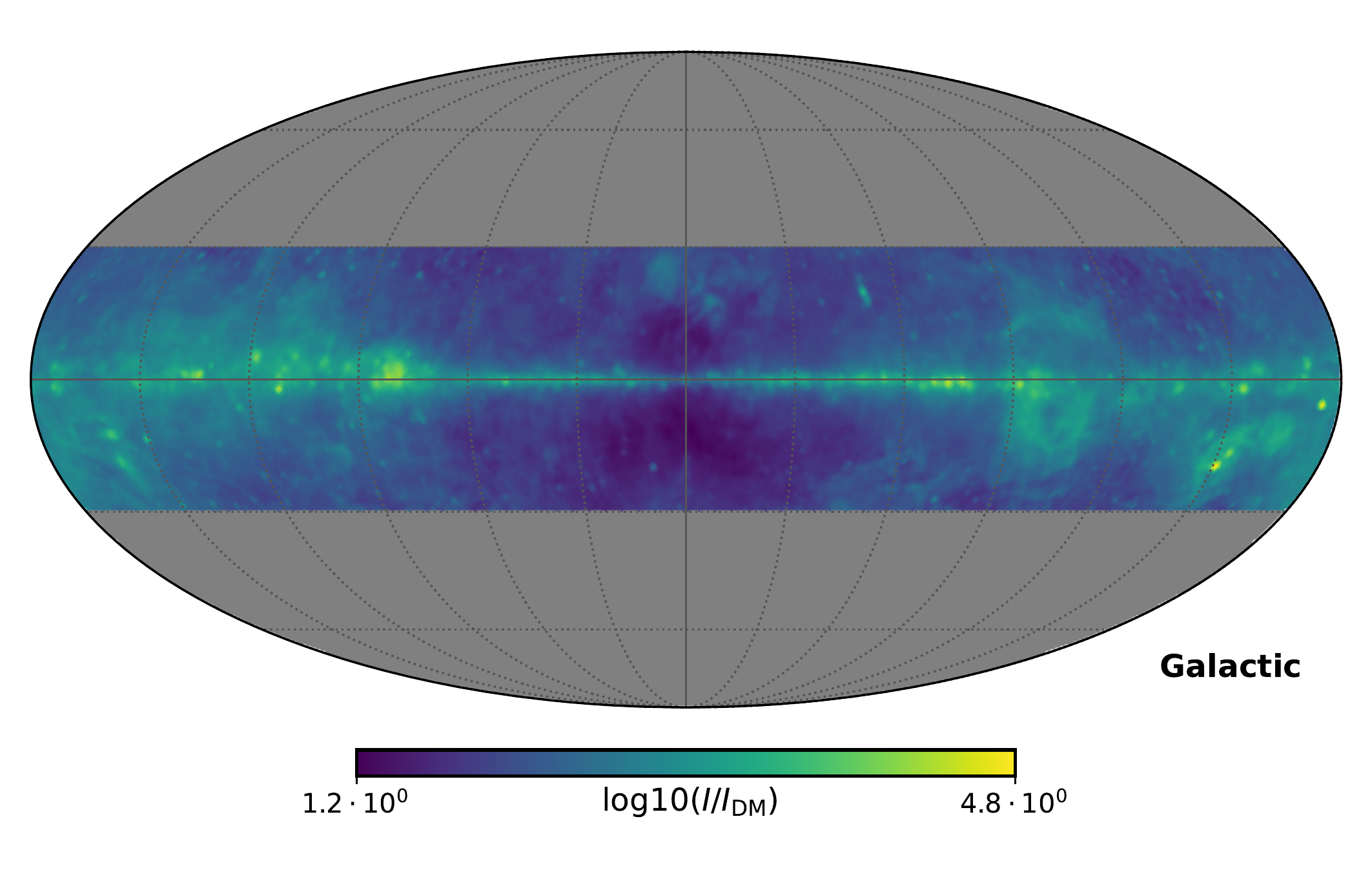}
		\includegraphics[width = 0.49\textwidth]{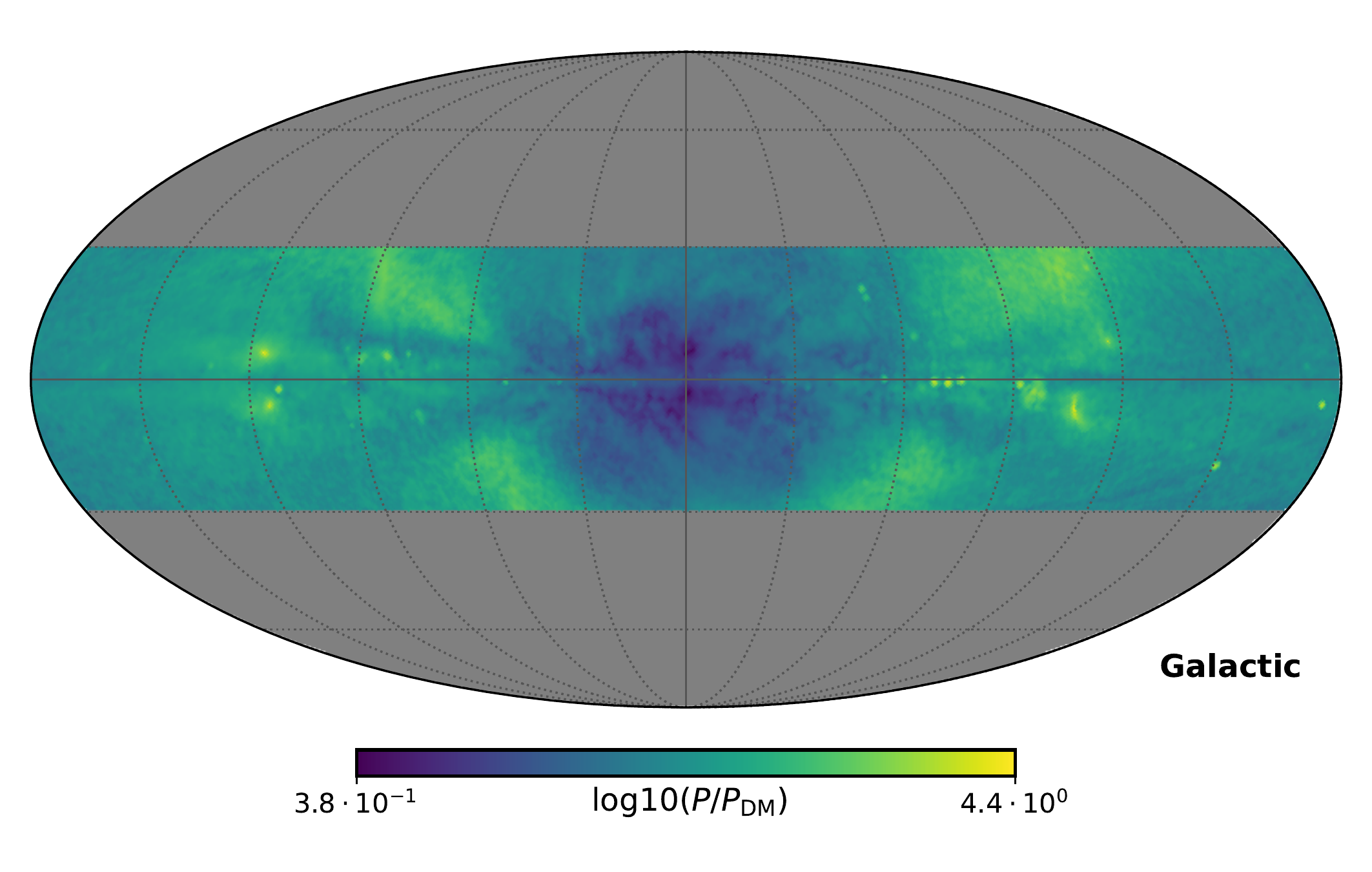}
	\caption{\textbf{Ratio between \planck data plus error estimation and synchrotron emission  from DM  at 30 GHz} (left: intensity, right: polarization) as computed for $m_{\rm DM}=50$~GeV annihilating into  $\mu^+ \mu^-$ pairs with a thermal averaged cross section of $\langle \sigma v\rangle = 3 \times 10^{-26}$ cm$^3$s$^{-1}$.
	 The propagation setup, GMF model and NSide resolution are the same as the ones used for Fig.~\ref{fig:mapIPdm}.}
	\label{fig:mapratio}
\end{figure*}

\begin{figure*}[t]
	\centering
	\includegraphics[width = 0.49\textwidth]{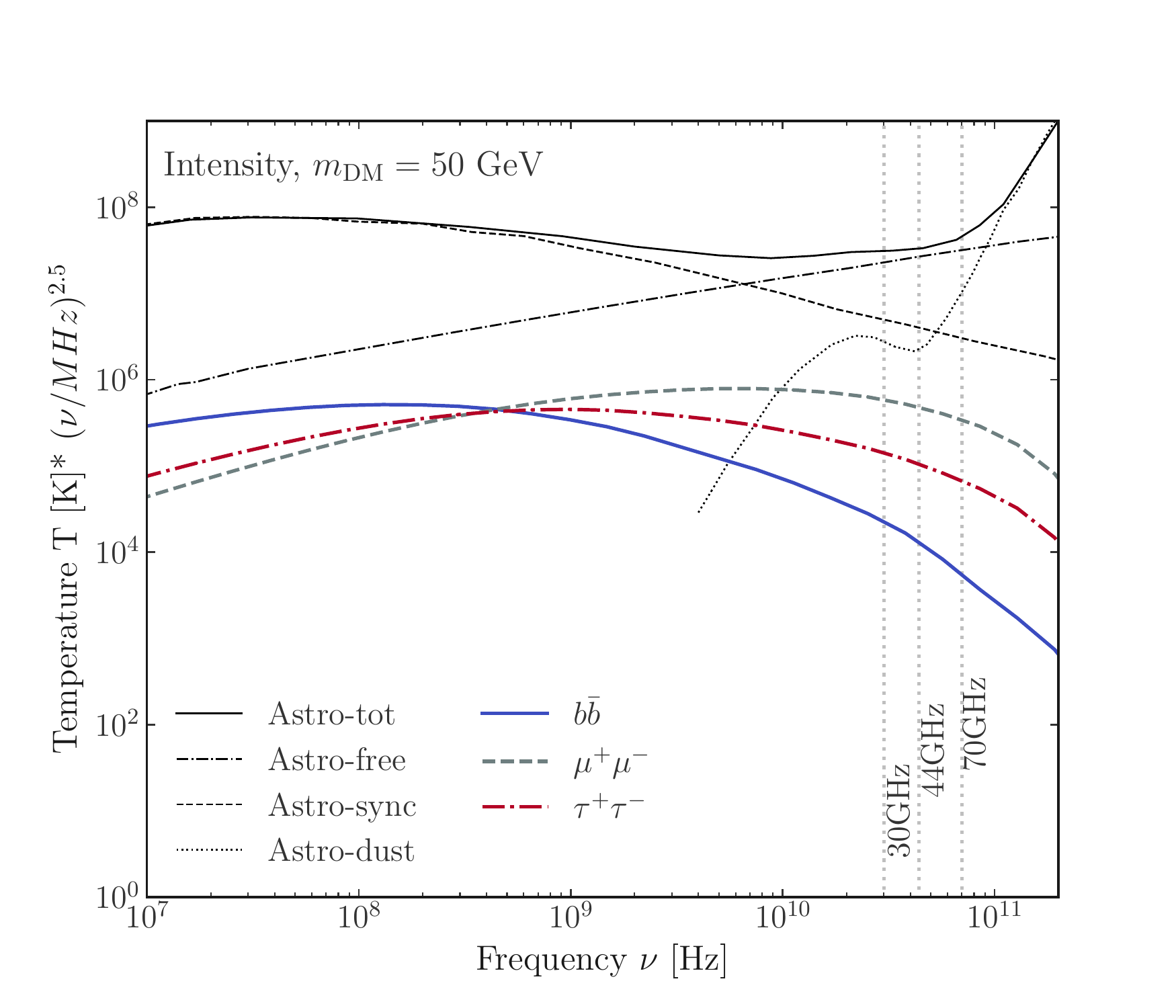}
	\includegraphics[width = 0.49\textwidth]{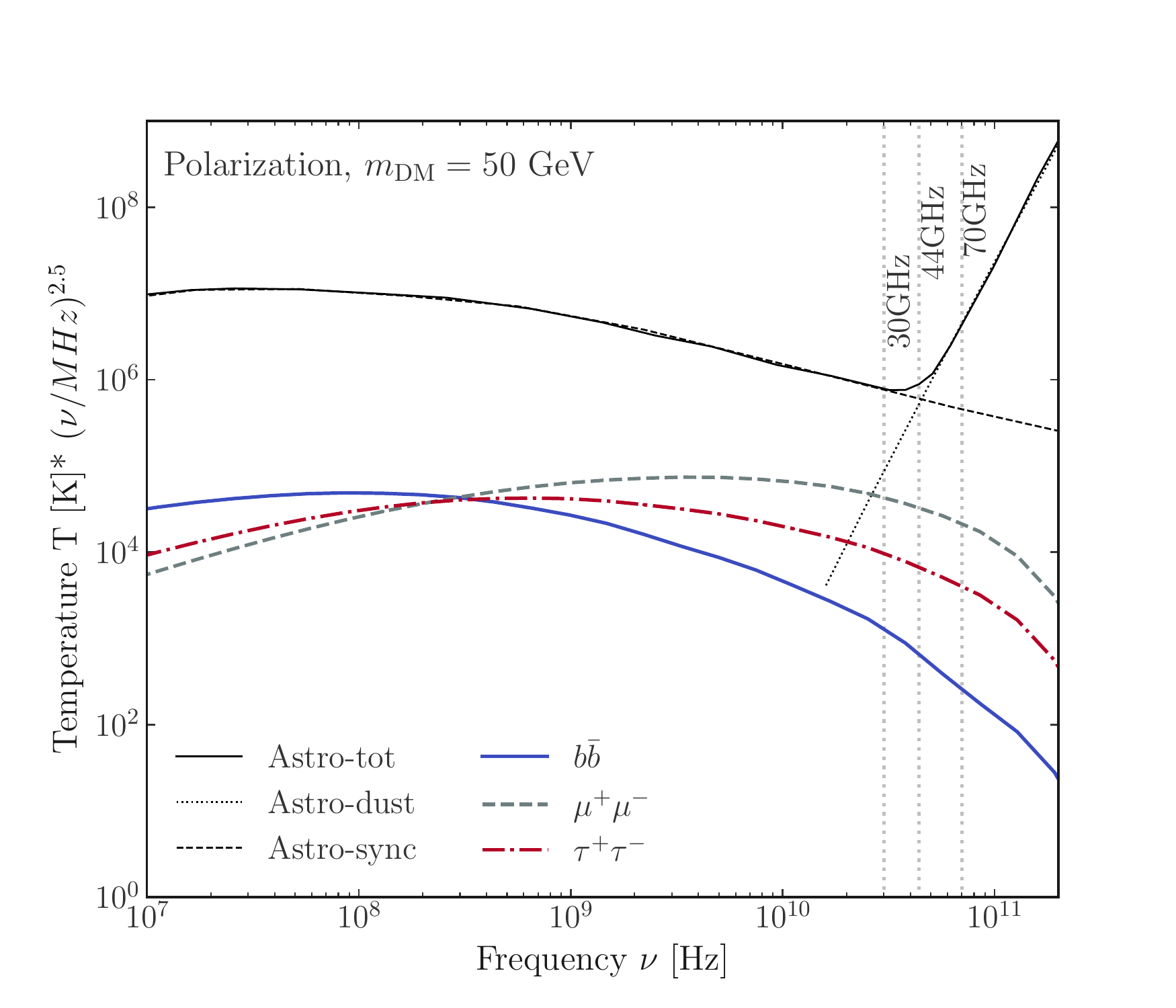}
	\\
	\includegraphics[width = 0.49\textwidth]{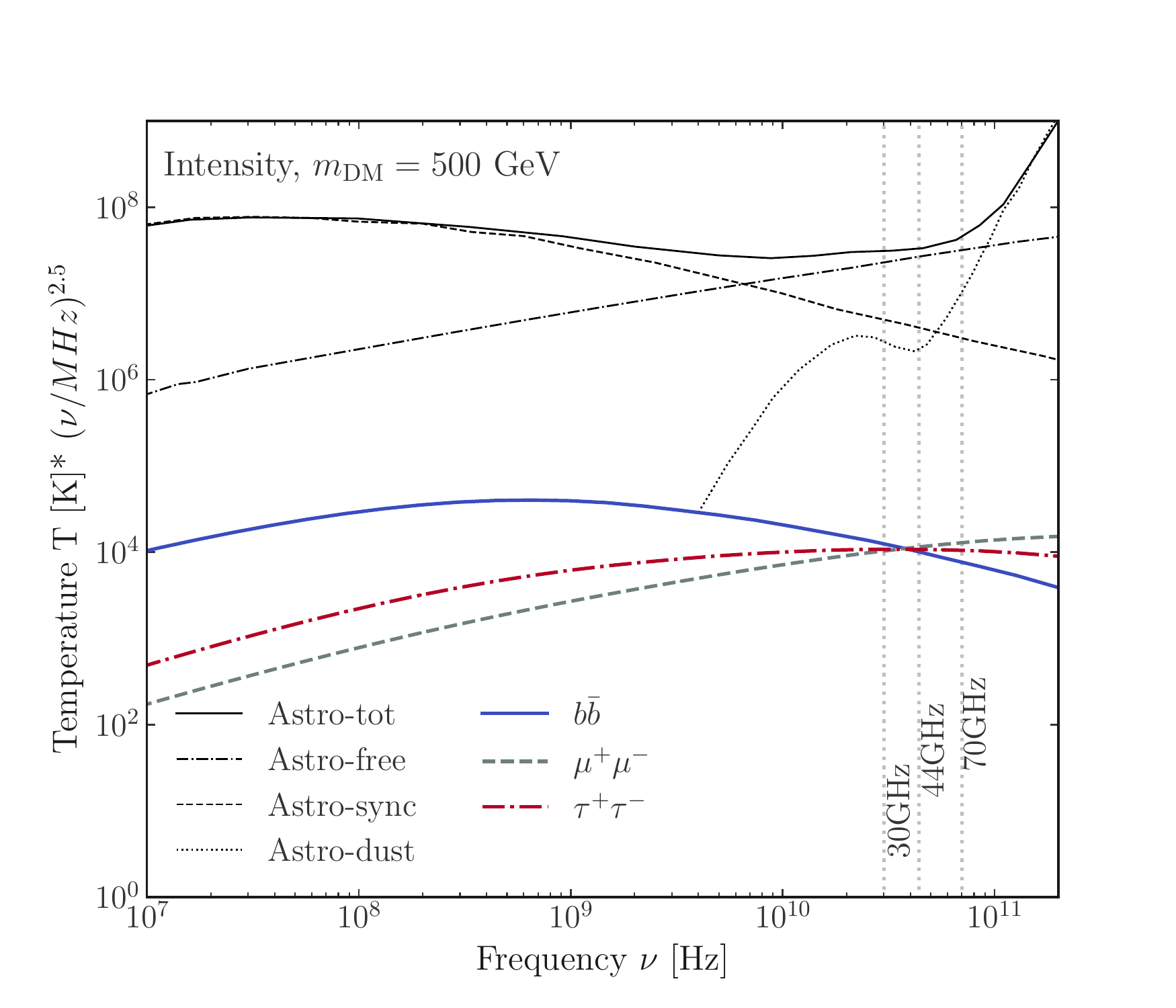}
	\includegraphics[width = 0.49\textwidth]{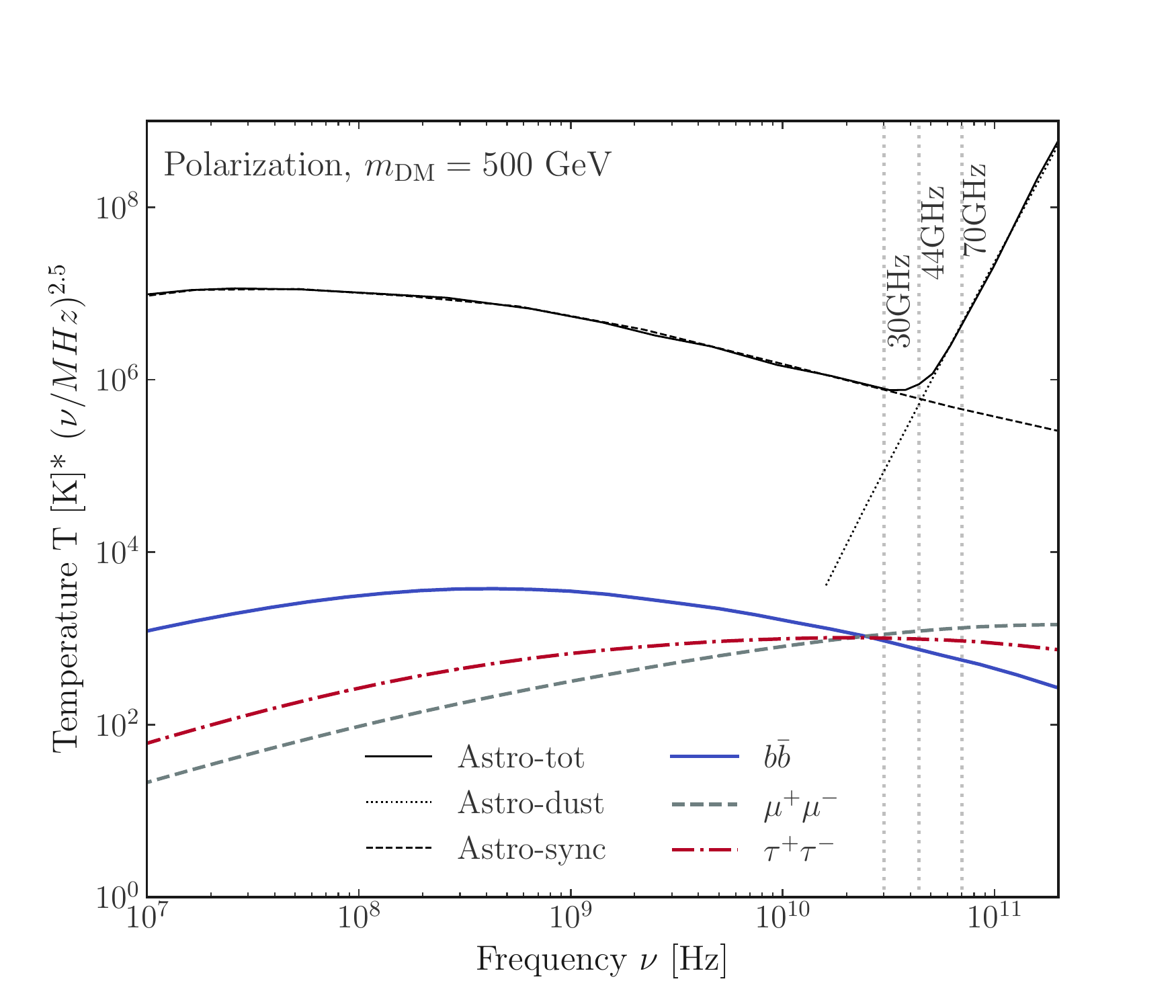}
	\caption{\textbf{Synchrotron emission spectrum from DM annihilations} for total intensity (left panels) and polarization amplitude (right panels) for different channels and DM mass of $50$~GeV (upper row) and $500$~GeV (lower row). Vertical dotted lines indicate the frequencies of \planck LFI. For all panels the DM signal is computed with \texttt{GALPROP} for the line of sight of (l,b)=(20,20)deg, and assuming PDDE propagation parameters, the Psh+11 GMF model and $\langle \sigma v\rangle = 3 \times 10^{-26}$~cm$^3$s$^{-1}$.   The 'Astro' spectra (total, free-free, synchrotron and dust) are taken from a representative model for the astrophysical intensity and polarization emission at high latitudes which fits multiwavelength data taken from  Ref.~\cite{Orlando:2013ysa}. }
	\label{fig:sedIPdm}
\end{figure*}


\section{Dark matter signal and constraints: extended results}\label{ssec:ul}

\subsection{Dark matter maps}
Fig.~\ref{fig:mapIPdm} shows the DM synchrotron maps for $30$~GHz for the  intensity (left) and polarization amplitude (right), and for different GMF models (one for each row). 
The morphologies of the Sun+10 and Psh+11 DM intensity signal (left panels, first two rows) are both peaked at the Galactic Center and  very similar, consistently with the fact that they use the same model for the random magnetic field, i.e., a double exponential.
The JF12 model (left panel, lower row) instead predicts a more structured DM signal, with a bright peak at the Galactic center and other peaks in the Galactic plane reflecting the random field structure in the spiral arms.
Overall, the JF12 model predicts a larger DM signal. This is explained by the larger field strength, see parameters in Ref.\cite{Jansson:2012rt} for more details. 
The morphology of the polarization amplitude signal is instead more complicated. The signal again peaks at the Galactic center, similarly to the intensity signal, but extends to higher latitudes following the ordered halo field.
Again, the JF12 model predicts a larger signal away from the Galactic center, given the presence of a X-field and striated component on top of the disk and halo fields. 

We show in Fig.~\ref{fig:mapratio} the map of the ratio between the \planck data at 30~GHz and a benchmark map for the annihilation of $m_{\rm DM}=50$~GeV into $\mu^+ \mu^-$ computed with the PDDE propagation and the Psh+11  model (so the same as the upper row in Fig.~\ref{fig:mapIPdm}). 
Before computing the ratio, the error estimate for intensity and polarization has been added to the respective \planck data map.
The lower value in the color map corresponds thus to our 68\% C.L. upper limit, see main text. 
Since the DM signal and the backgrounds have different morphologies, the most constraining region does not coincide with the Galactic center, although the two signals both peak at the Galactic center.
Instead, the most constraining pixels  are located in a region various degrees above or below the Galactic Plane where the background is lower and the DM signal is still significant (compare with Fig.~\ref{fig:mapIPdm}), providing the optimal signal to noise (S/N). 
For the polarization the optimal region is closer to the Galactic plane (GLON 358.2, GLAT 6.6 degrees) than in the case of the intensity (GLON 356.5, GLAT -14.8 degrees). This is due to the filamentary morphology of the polarization Planck map which leaves regions of low background in between the filaments very close to the Galactic center, contrary to the intensity case. This in part explains why  polarization is more constraining than intensity regarding the DM signal.

Since we do not mask the microwave point sources in the map, their positions correspond to large values of log10($I/I_{\rm  DM}$), and log10($P/P_{\rm  DM}$), and thus weak DM constraints.

\subsection{Dark matter spectra}
To illustrate the synchrotron spectra we explore a wide range of frequencies above and below the \planck LFI values in Fig.~\ref{fig:sedIPdm}, for fixed masses of $m_{\rm DM}=50,500$~GeV (left and right panels, respectively) and the three benchmark annihilation channels. 
The spectrum of the polarization amplitude is very similar to the intensity one, with an overall smaller normalization value at the representative line of sight of (l,b)=(20,20)deg, at intermediate latitudes in the sky.
At \planck LFI frequencies, an higher signal is expected in the leptonic channels, in particular in the $\mu^+ \mu^-$. 
The hadronic channels might be better constrained using data at lower frequencies, see e.g. Ref.~\cite{Cirelli:2016mrc}.

For a tentative comparison, we include the spectra of the background emissions ('Astro') as estimated in Ref.~\cite{Orlando:2013ysa} at high latitudes.  Specifically, we take the total astrophysical emission which fits their multiwavelength dataset at high latitudes (right hand plots in their Figure 4), as well as the individual contributions estimated for synchrotron, dust, and free-free emission. For polarization below few tens of GHz, this is just the synchrotron contribution; for intensity, at the WMAP/\planck frequencies the free-free and spinning dust contributions are also important. 
As a further argument compelling our main results, we see that  at the \planck frequencies the DM polarization signal is a factor 3-4 closer to the astrophysical emission derived in Ref.~\cite{Orlando:2013ysa} (which fits the data) with respect to the DM intensity. 
A  computation of the astrophysical synchrotron intensity and polarization emission maps and spectra within the same sky region, propagation and GMF model is left to future work.

\subsection{Pixel size}
\begin{figure}[t]
	\centering
	\includegraphics[width = 0.5\textwidth]{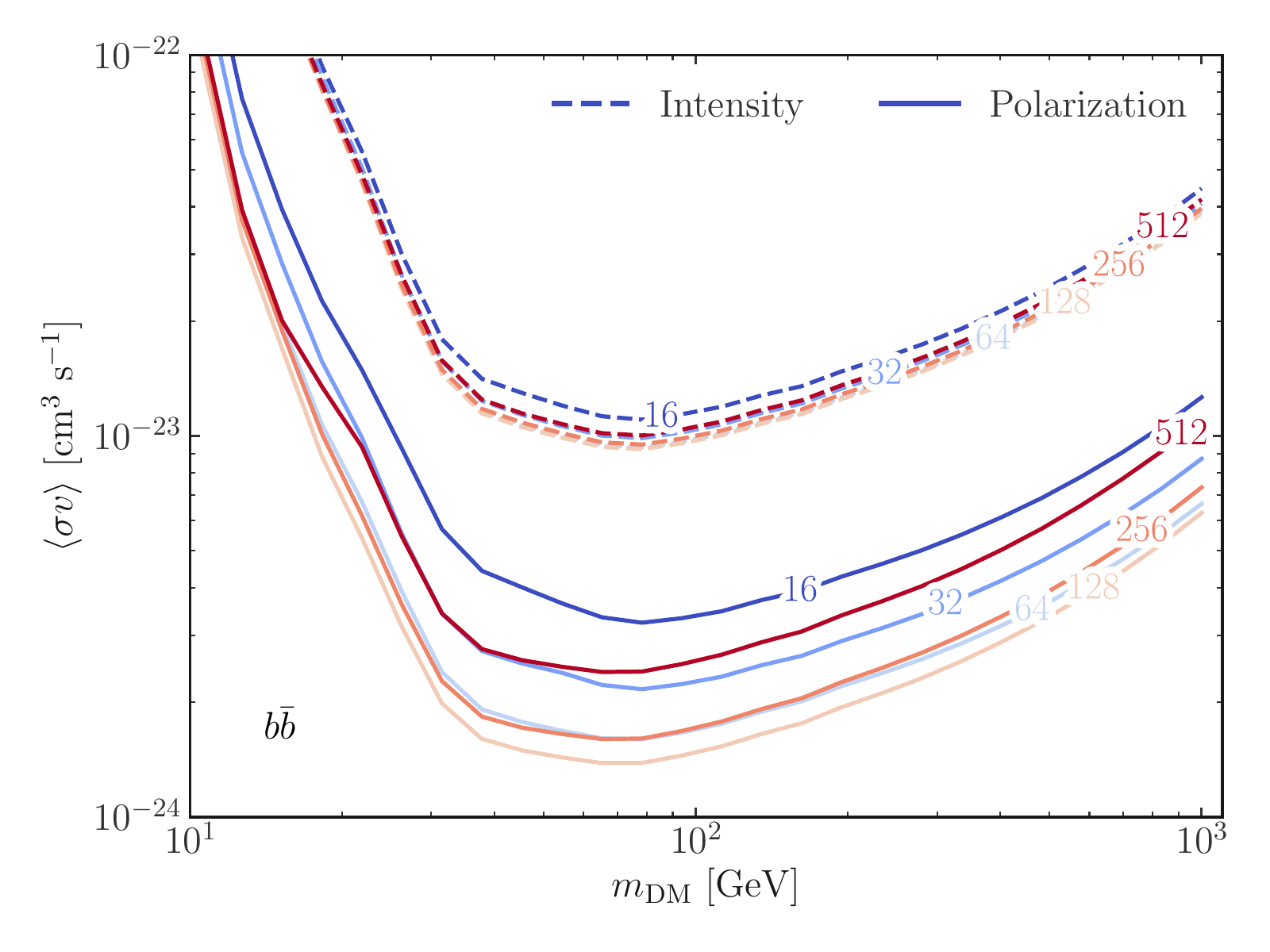}
	\caption{\textbf{Effect of the pixel size} on the upper limits on the thermally averaged annihilation cross section as a function of DM mass as derived from the \planck intensity (dashed lines) and polarization (solid lines) data at $30$~GHz. The color scale, together with the line label indicate the \healpix resolution NSide, from 16 (blue) to 512 (red). }
	\label{fig:pixeffect}
\end{figure}
To investigate the effect of the pixel size in deriving DM upper limits with the \planck intensity and polarization maps we compute the constraints for different NSide values, from 16 to 512. 
For each resolution, the data and error maps are built as described in the main text, while the \galprop prediction computed for NSide=512 is downgraded to low resolution using the \texttt{healpy.ud\_grade} routine. 
We show the resulting upper limits for the synchrotron intensity (dashed lines) and the polarization amplitude (solid lines) as a function of the NSide value (line label,  color scale from blue (16) to red (512)) in Fig.~\ref{fig:pixeffect}. 
The results are obtained fixing the PDDE propagation parameters, the Psh+11 GMF model and the $\bar{b} b $ annihilation channel. 
We first note that the DM constraints from intensity are only weakly dependent on the pixel size. This is consistent with the fact that the \planck measured intensity is in a regime of high S/N, as indeed can be seen comparing the intensity and error maps in Fig.~\ref{fig:planck}.
On the contrary \planck polarization has still a large noise and the pixel size has a larger impact on the DM constraints. A large pixel size increase the S/N at the price of losing the details of the morphology, while with a small pixel size the morphology of the signal is retained but with a lower S/N. As can be seen in Fig.~\ref{fig:pixeffect}, indeed, Nside 16 and NSide 512 provide the worst constraints while the optimal constraints are provided by the intermediate choice of NSide 128, i.e., for pixel mean spacing of about 0.5 deg, which is the one adopted for the results in the main text.
We note, nonetheless, that even for the worst cases of Nside 16 and NSide 512 the polarization constraints are better than the intensity ones by a factor of 5, while in the optimal case of NSide 128 they are better by about one order of magnitude. 

\subsection{Dark matter density profile}
\begin{figure*}[t]
	\centering
		\includegraphics[width = 0.49\textwidth]{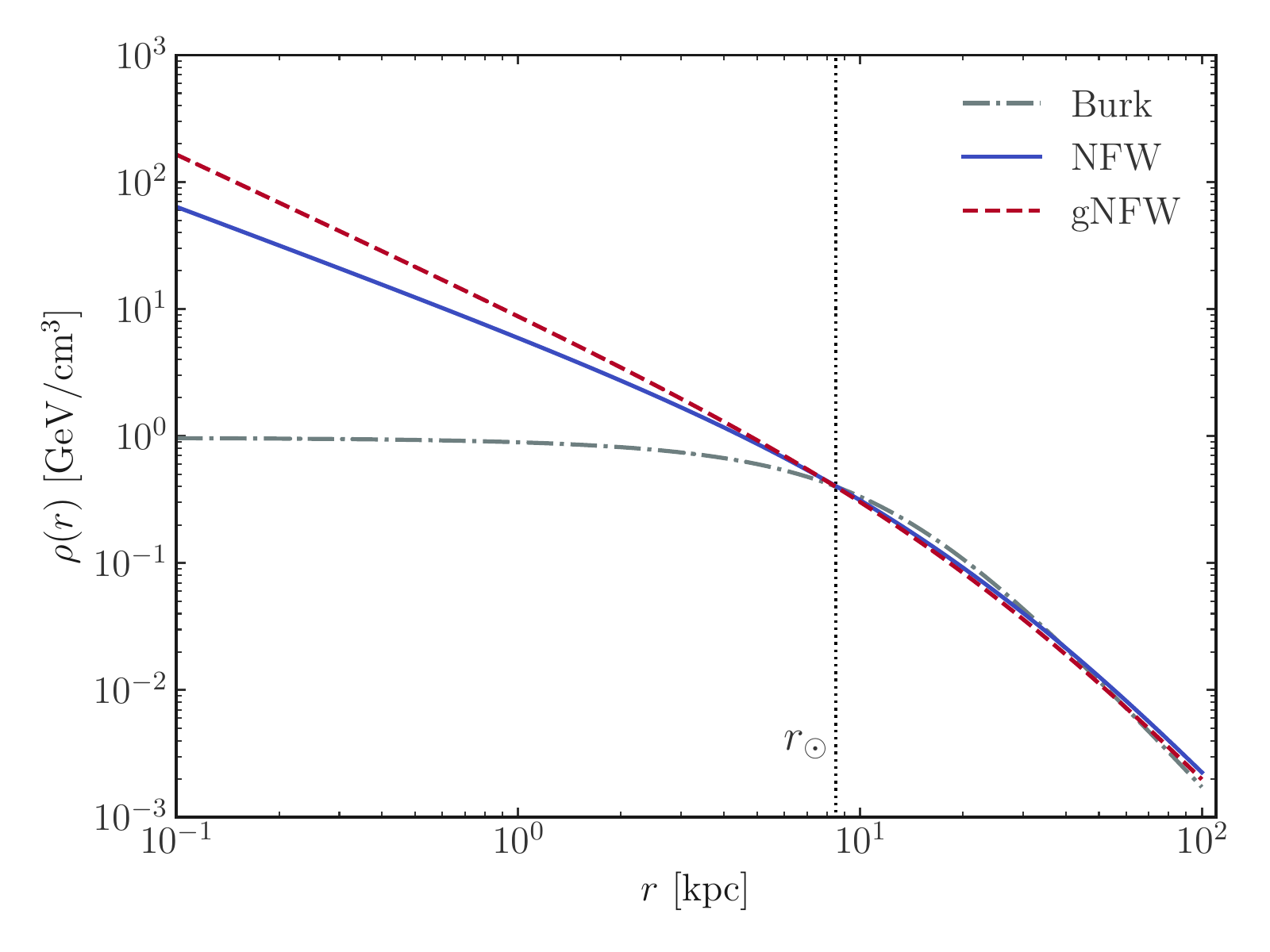}
				\includegraphics[width = 0.49\textwidth]{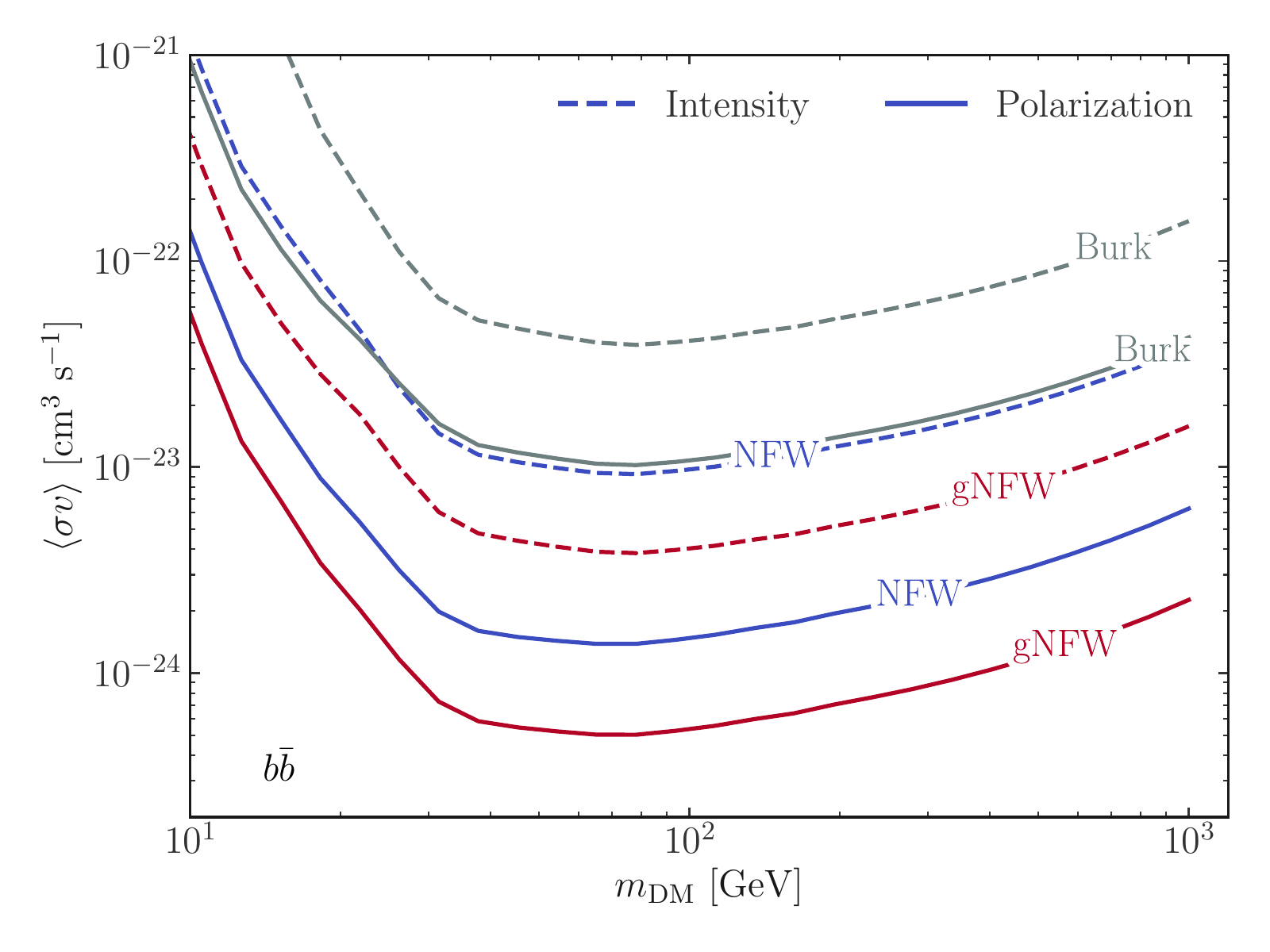}
	\caption{\textbf{Left: DM radial density profiles} as a function of the distance from the Galactic center $r$ used in this work, the benchmark being the NFW profile (blue solid line). The position of the Earth is indicated by a dotted line. The local DM density is set to be $\rho_{\rm DM}(r_\odot =8.5~\rm{kpc})=0.4$~GeV/cm$^3$. \textbf{Right: effect of the DM density profile} on the upper limits on the thermally averaged annihilation cross section as a function of the DM mass, as derived from \planck intensity (dashed lines) and polarization (solid lines) data at $30$~GHz and for $\bar{b} b$ DM annihilation.}
	\label{fig:DMprofile}
\end{figure*}

The effect of varying the DM density profile on the upper limits is illustrated in Fig.~\ref{fig:DMprofile} (right panel). 
The blue lines refer to the benchmark NFW model, while the red and gray lines to the gNFW and Burkert cored profile, respectively. 
As it can be seen, the systematic uncertainty connected to the choice of the DM density profile is about one order of magnitude for both intensity and polarization. 
This is expected since the constraining power of our analysis comes from the Galactic center region (compare with Fig.~\ref{fig:mapratio}), where the DM signal peaks, and in this region the different choices of profile differ significantly (see Fig.~\ref{fig:DMprofile} left panel) giving large differences in the predicted DM signal.
Our benchmark choice, the NFW profile, provides an intermediate result between the gNFW and the Burkert profile.

\subsection{Propagation parameters}
\begin{figure*}[t]
	\centering
		\includegraphics[width = 0.49\textwidth]{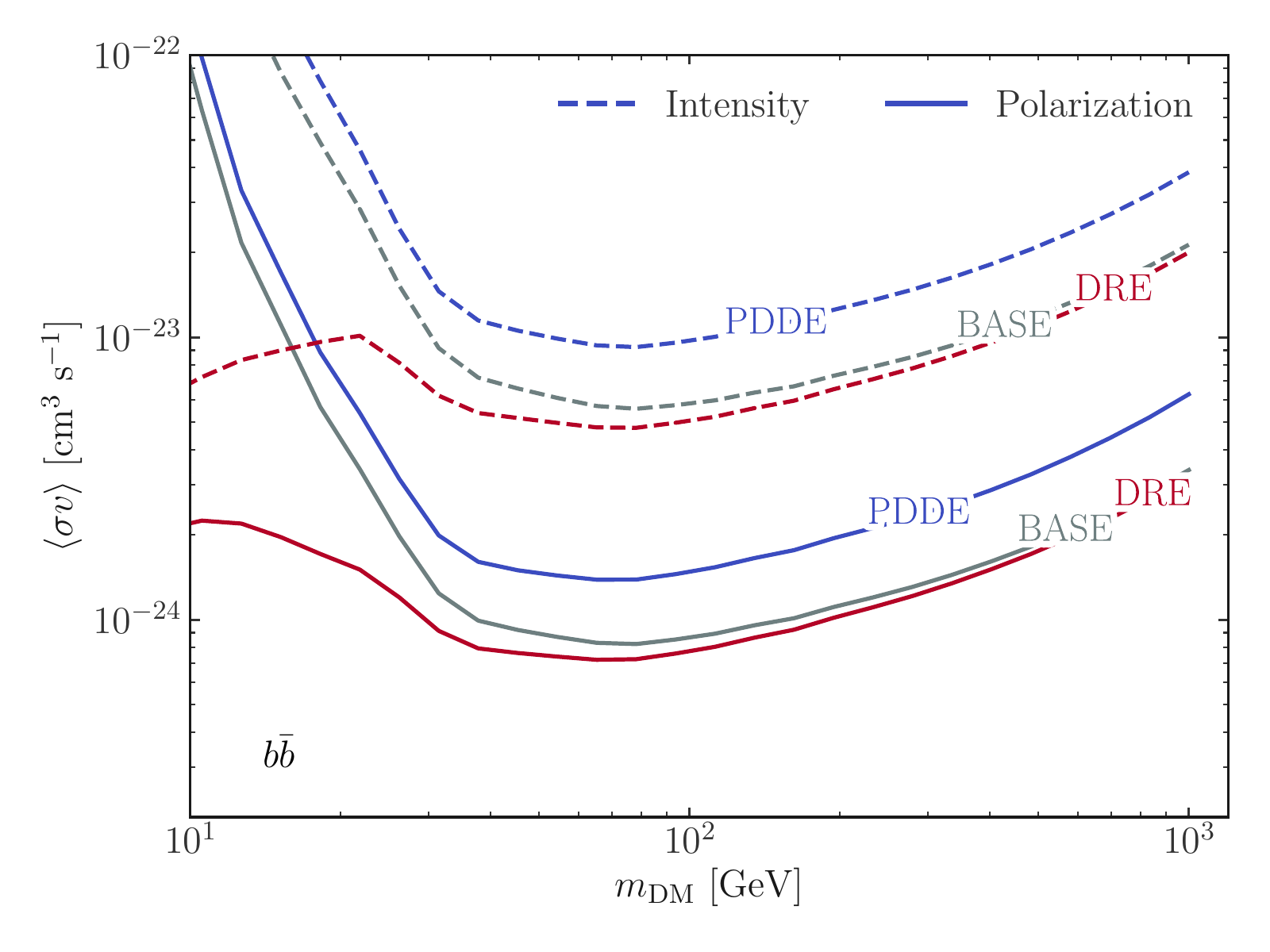}
			\includegraphics[width = 0.49\textwidth]{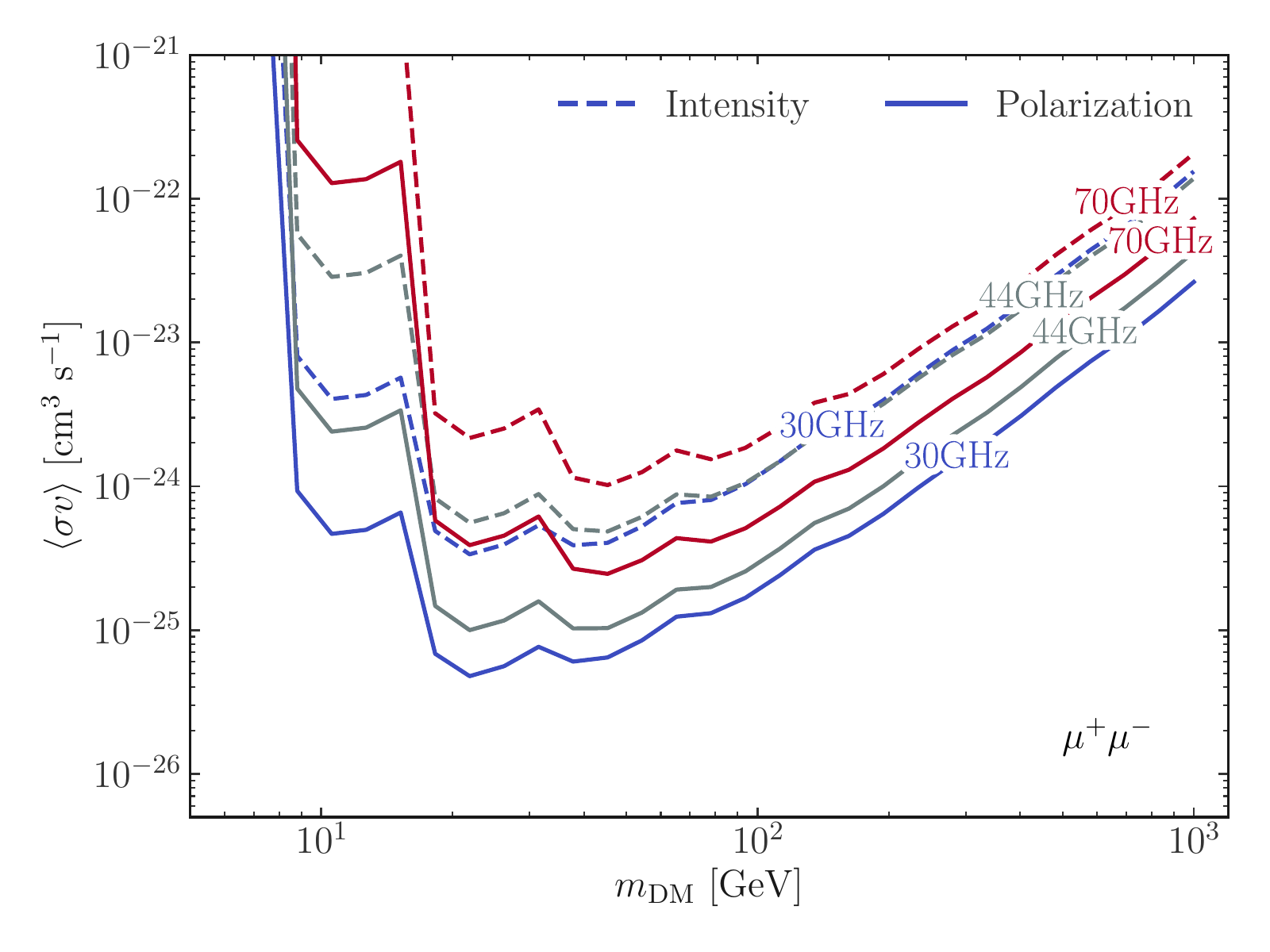}
	\caption{\textbf{Left: Propagation model effect} on the upper limits on the thermally averaged annihilation cross section into $\bar{b} b$ pairs as a function of DM mass as derived from the \planck intensity (dashed lines) and polarization (solid lines) data at $30$~GHz.
	\textbf{Right: Results for all the  \planck LFI frequencies} for the $\mu^+ \mu^-$ channel.  
	}
	\label{fig:propfreq}
\end{figure*}

Another important systematic uncertainty for our DM upper limits is connected to the choice of the propagation setup. We illustrate the results for the three models we have employed in Fig.~\ref{fig:propfreq} (left).
It can be seen that the limits differ in particular for low DM masses. 
Specifically, the DRE model, which is the only one with non-vanishing reacceleration, provides more stringent constraints at masses $m_{\rm DM}<30$~GeV. This is because the rather high Alfven velocity in the DRE model increases the density of low energy $e^\pm$, and so their synchrotron emission, see Refs.~\cite{2018MNRAS.475.2724O,2019PhRvD..99d3007O} for the corresponding propagated $e^\pm$ spectra at Earth. 
The PDDE propagation model provides the most conservative results among the explored models. 
Overall, the systematic uncertainty related to propagation effects is at the level of 20-30\% for DM masses above 30 GeV.

\subsection{Upper limits for 44-70 GHz}
The upper limits computed using all the three \planck LFI frequency maps are illustrated in Fig.~\ref{fig:propfreq} (right panel) for the case of DM annihilating into $\mu^+ \mu^-$ pairs.
The limits for 44 and 70 GHz are obtained following the same procedure as the $30$~GHz case. 
We find that the data at $30$~GHz provide the most constraining results in all the DM mass interval, with the only exception being a slight improvement for the $44$~GHz intensity case at masses larger than about 100~GeV. 
In principle, looking at the synchrotron emission spectrum in Fig.~\ref{fig:sedIPdm},
an improvement of the limits at high DM masses for the leptonic annihilation channels is expected, especially for polarization, since the ratio of signal over background is increasing with increasing frequency, confront, e.g, the lower panels of Fig.~\ref{fig:sedIPdm} for the case $m_{\rm DM}=500$~GeV.
In practice, however, the relative error on the \planck measured polarization increases significantly at 44 and 70 GHz (see maps in Figs.\ref{fig:planck44}-\ref{fig:planck70}) and this degrades the constraints at a level that makes the previous expectation not satisfied.
We verified, indeed, that, when not considering the error estimation in the upper limit computation, the $44,70$~GHz frequencies are more constraining for $m_{\rm DM}\gtrapprox 100$~GeV.

\subsection{Comparison with other works and probes}

\begin{figure*}[t!!]
	\centering
		\includegraphics[width = 0.49\textwidth]{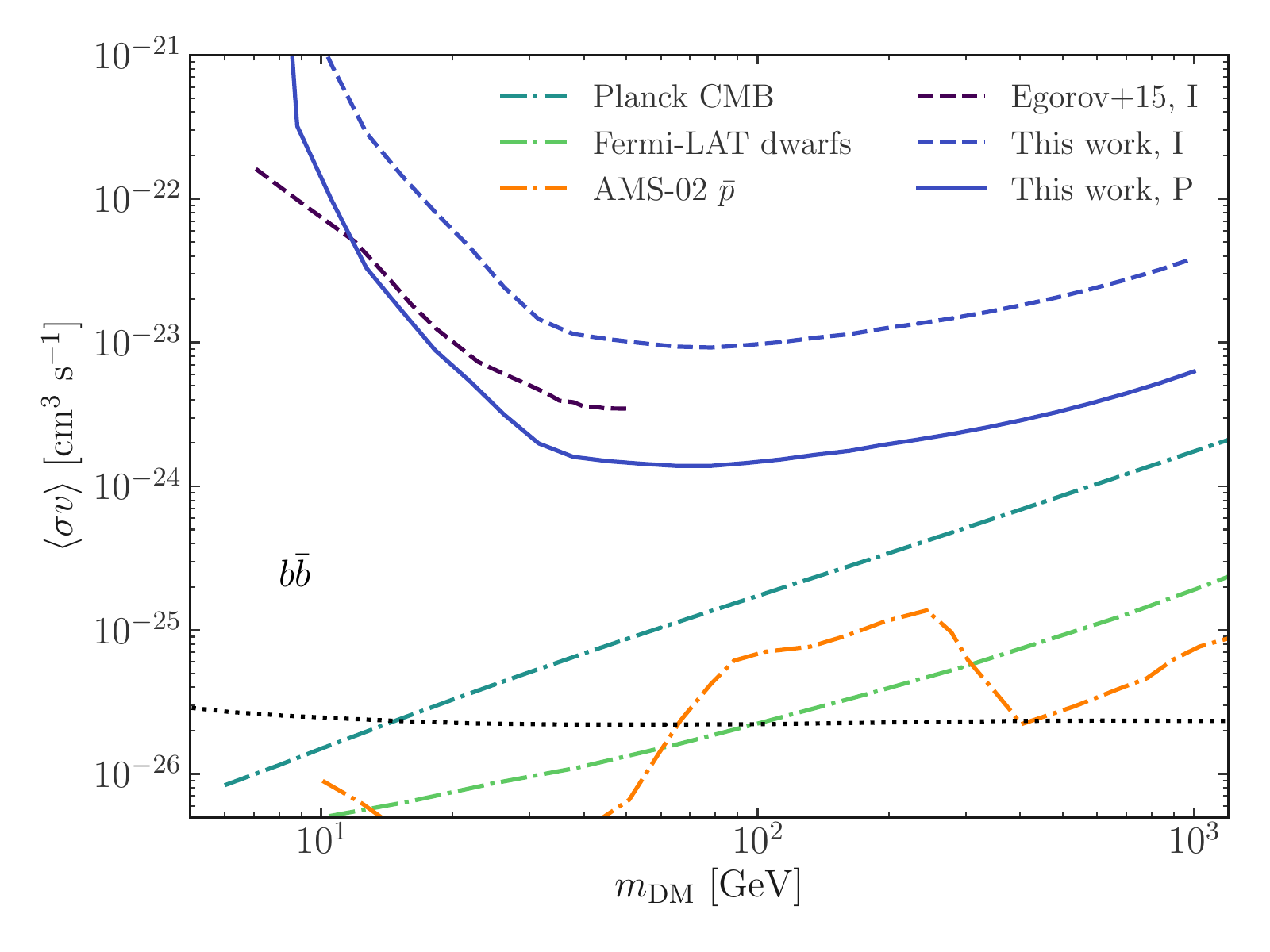}
			\includegraphics[width = 0.49\textwidth]{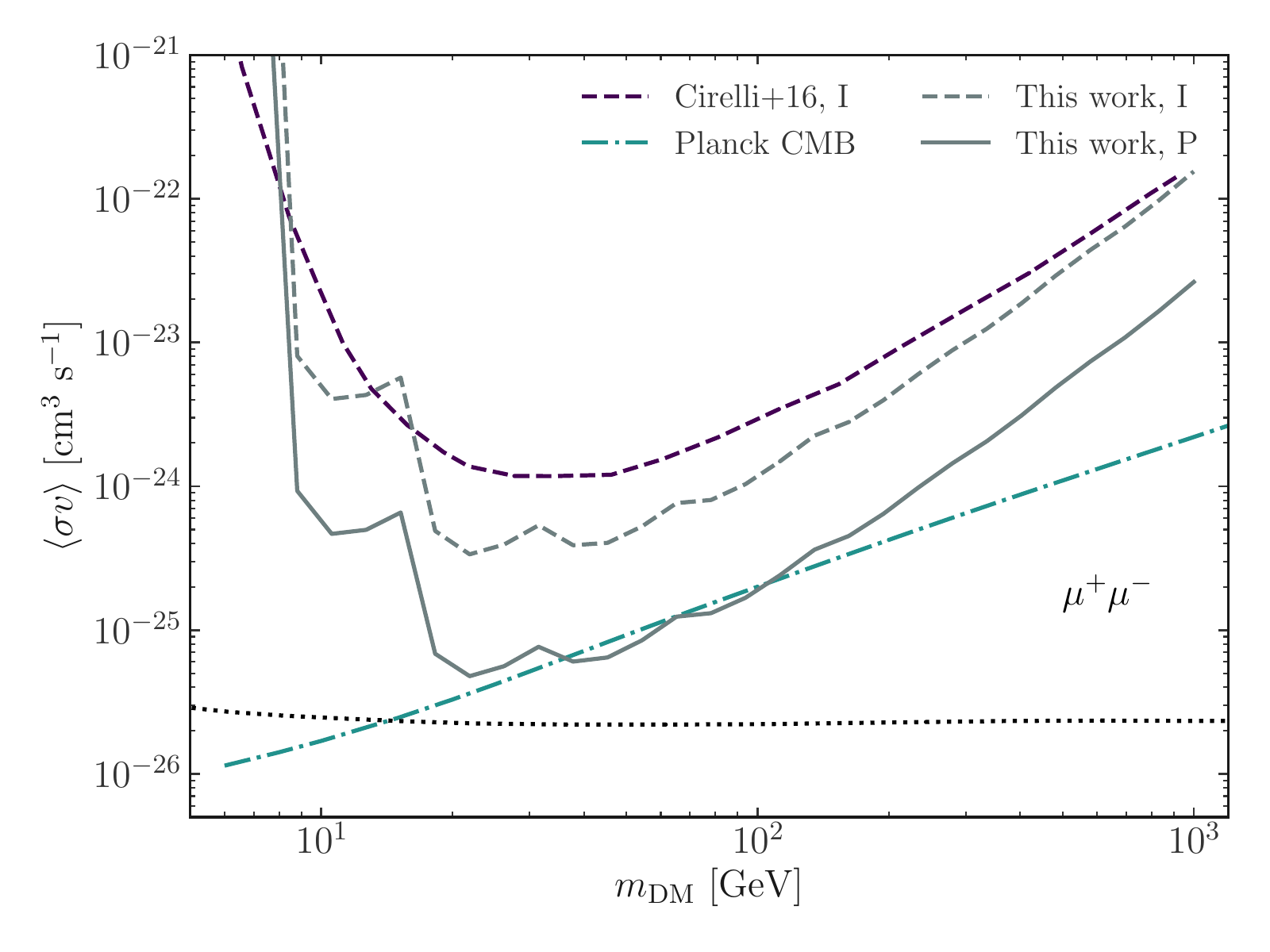}
	\caption{\textbf{Comparison between our conservative constraints and the ones from other works and other probes.} Our constraints are derived from the \planck intensity (blue dashed lines) and polarization (blue solid lines) data at $30$~GHz using benchmark Psh+11 GMF model and PDDE propagation parameters.
	 Left: Results for $\bar{b} b$ channel are compared to the analysis of ref.~\cite{Egorov:2015eta} (which uses WMAP and \planck data and synchrotron intensity only), CMB limits obtained from \planck data \cite{Planck:2018vyg}, limits obtained from gamma-ray observations of dwarfs with Fermi-LAT \cite{Fermi-LAT:2016uux} and from  AMS-02 CR $\bar{p}$~\cite{Kahlhoefer:2021sha}.
	Right:  Results for $\mu^+ \mu^-$ channel are compared with 
	the analysis of ref.~\cite{Cirelli:2016mrc} of early \planck data at 30~GHz using synchrotron intensity only and CMB limits obtained from \planck data~\cite{Planck:2018vyg}.
	The horizontal dotted line indicates the thermal relic annihilation cross section expectation \cite{Steigman:2012nb}. }
	\label{fig:compare}
\end{figure*}

We show in Fig.~\ref{fig:compare} a comparison of our benchmark results (Psh+11 GMF, PDDE propagation, $30$~GHz \planck data) with representative upper limits obtained with similar or complementary probes of DM annihilation in our Galaxy and beyond. 
By exploiting the DM Galactic signal from synchrotron intensity only and early microwave data from WMAP and \planck, the authors of Ref.~\cite{Egorov:2015eta} (left panel, $\bar{b} b$ channel) and Ref.~\cite{Cirelli:2016mrc} (right panel, $\mu^+ \mu^-$ channel) obtain results similar to our intensity constraints.
Differences can be explained in terms of different choices of
the GMF, propagation parameters and DM density profile, see the respective papers for more details.
Results for the $\bar{b} b$ channel are not competitive with other DM targets and probes, such as traditional template fitting analysis of  gamma-rays from dwarfs in Fermi-LAT  data \cite{Fermi-LAT:2016uux} (see also more conservative, data-driven results presented in Ref.~\cite{Calore:2018sdx}) or AMS-02 $\bar{p}$ (limits taken from Ref.~\cite{Kahlhoefer:2021sha}, comparable to other works, see e.g., \cite{Cuoco:2016eej,2022arXiv220203076C}). This is expected since the synchrotron signal probes mostly the leptonic annihilation channels.
Indeed, our results using \planck polarization are competitive with \planck CMB constraints~\cite{Planck:2018vyg} between about 50~GeV and 100~GeV for the $\mu^+ \mu^-$ channel. This  motivates further interest in going beyond the conservative approach presented in this paper.

\end{document}